\documentclass[apj]{emulateapj}

\usepackage{graphicx} 
\usepackage{amsmath}
\usepackage{hyperref}
\usepackage{amsfonts}
\usepackage{amsmath}
\usepackage{amssymb}
\usepackage{amsthm}
\usepackage{subeqnarray}
\usepackage{ulem}

\newcommand{\delad}{\nabla_{\rm ad}}
\newcommand{\delrad}{\nabla_{\rm rad}}

\DeclareMathSymbol{\varOmega}{\mathord}{letters}{"0A}
\DeclareMathSymbol{\varSigma}{\mathord}{letters}{"06}
\DeclareMathSymbol{\varPsi}{\mathord}{letters}{"09}

\newcommand{\Eq}[1]{Equation\,(\ref{#1})}

\newcommand{\App}[1]{Appendix~\ref{#1}}

\newcommand{\RB}{R_{\rm B}}
\newcommand{\co}{_{\rm c}}
\newcommand{\di}{_{\rm d}}
\newcommand{\cb}{_{\rm RCB}}

\begin{document}
\bibliographystyle{apj}

\shortauthors{Piso, Youdin \& Murray-Clay}

\title{Minimum Core Masses for Giant Planet Formation With Realistic Equations of State and Opacities}
\author{Ana-Maria A. Piso\altaffilmark{1}, Andrew N. Youdin\altaffilmark{2}, Ruth A. Murray-Clay\altaffilmark{1,3}}
\altaffiltext{1}{Harvard-Smithsonian Center for Astrophysics, 60 Garden Street, Cambridge, MA 02138; apiso@cfa.harvard.edu}
\altaffiltext{2}{Steward Observatory, University of Arizona, 933 N Cherry Ave, Tucson AZ 85721}
\altaffiltext{3}{Department of Physics, University of California, Santa Barbara, CA 93106}

\begin{abstract}

Giant planet formation by core accretion requires a core that is sufficiently massive to trigger runaway gas accretion in less that the typical lifetime of protoplanetary disks. We explore how the minimum required core mass, $M_{\rm crit}$, depends on a non-ideal equation of state and on opacity changes due to grain growth, across a range of stellocentric distances from 5-100 AU. This minimum $M_{\rm crit}$ applies when planetesimal accretion does not substantially heat the atmosphere.  Compared to an ideal gas polytrope, the inclusion of molecular hydrogen (H$_2$) dissociation and variable occupation of H$_2$ rotational states increases $M_{\rm crit}$.  Specifically, $M_{\rm crit}$ increases by a factor of $\sim$$2$ if the H$_2$ spin isomers, ortho- and parahydrogen, are in thermal equilibrium, and by a factor of $\sim$$2-4$ if the ortho-to-para ratio is fixed at 3:1.  Lower opacities due to grain growth reduce $M_{\rm crit}$. For a standard disk model around a Solar mass star, we calculate $M_{\rm crit} \sim 8 M_{\oplus}$ at 5 AU, decreasing to $\sim$$5 M_{\oplus}$ at 100 AU, for a realistic EOS with an equilibrium ortho-to-para ratio and for grain growth to cm-sizes. If grain coagulation is taken into account, $M_{\rm crit}$ may further reduce by up to one order of magnitude. These results for the minimum critical core mass are useful for the interpretation of surveys that find exoplanets at a range of orbital distances.

\end{abstract}

\section{Introduction}
\label{intro}

Core accretion --- a prominent theory of giant planet formation --- stipulates that Jupiter-sized planets form when planetesimal accretion produces a solid core large enough to attract a massive atmosphere   (e.g., \citealt{mizuno78}, \citealt{stevenson82}, \citealt{boden86}, \citealt{wuchterl93}, \citealt{dangelo11}). Because protoplanetary disks dissipate on timescales of a few Myrs \citep[e.g.,][]{jay99}, cores must grow quickly to accrete disk gas. Fast  growth is particularly challenging far from a core's host star, where dynamical times are long. %, making the production of massive cores within the disk lifetime less likely.
Understanding whether core accretion can work in the outer disk may provide valuable insights into the formation of wide-separation planets such as the directly-imaged giants orbiting HR 8799 \citep{marois08}.

Production of such planets \textit{in situ} by core accretion requires faster core growth rates than typically assumed.  However, even if fast  growth could be achieved, it produces an additional difficulty.  Though the minimum, or critical, core mass to form a giant planet is often quoted as $M_{\rm crit}\sim 10M_{\oplus}$, the actual value depends on how quickly the core accretes planetesimals  (e.g.,  \citealt{pollack96}, \citealt{ikoma00}, \citealt{rafikov06}).  Fast accretion heats a core's atmosphere, increasing pressure support and hence $M_{\rm crit}$.  \citet{rafikov11} finds that beyond 40-50 AU, a core cannot  reach $M_{\rm crit}$ while accreting at the rate it requires to grow during its host disk's lifetime.

%Given standard core accretion rates, the $M_{\rm crit}$ you need to grow an atmosphere during the disk lifetime while the core is growing is too large, leading people to believe that core accretion doesn't work in the outer disk. 

This limit on the distance at which core accretion operates may be overcome if core growth does not proceed at a constant rate. Time-dependent models \citep[e.g.,][]{pollack96,ikoma00} suggest that a planets's feeding zone may be depleted of planetesimals before disk dissipation, causing core growth to stall. Planetesimal accretion no longer deposits energy into the atmosphere which, without this balance for radiative losses, cannot maintain a steady state.  The envelope accretes gas while  undergoing Kelvin-Helmholtz (KH) contraction.  In this regime, a core of any mass would evolve into a giant planet if given an infinite amount of time.  In practice, a core produces a giant planet if it has enough time to accumulate an atmosphere of approximately its own mass before its host disk dissipates.  The minimum core mass satisfying this criterion is $M_{\rm crit}$.

Because extra heating increases an atmosphere's pressure, opposing runaway envelope growth,  $M_{\rm crit}$ is smallest in the absence of planetesimal accretion.
 While $M_{\rm crit}$ has been systematically computed as a function of disk properties and stellocentric separation for steady-state atmospheres heated by planetesimal accretion \citep{rafikov06}, no equivalent systematic study is available for this minimum value of $M_{\rm crit}$.

In \citet[hereafter Paper I]{piso14} and this paper, we provide such a systematic study.   Given a fiducial disk model, we calculate the $M_{\rm crit}$ required to nucleate runaway atmospheric growth for a fully formed---and no longer accreting---core.  We model the planet's evolution using a series of quasi-static two-layer atmospheres embedded in a protoplanetary disk. Here, we build on the results of Paper I by making two important additions: (1) a realistic equation of state (EOS), and (2) realistic dust opacities.  Our aim is twofold: (1) to explain how hydrogen and helium's non-ideal EOS affects atmospheric growth when compared to an ideal gas, and (2) to provide realistic estimates for the minimum critical core mass required to form a giant planet over a range of semimajor axes.

\pagebreak

\subsection{Equation of State}
\label{sec:introeos}

We use the EOS of hydrogen and helium mixtures calculated by \citet{saumon95}, which captures non-ideal effects such as dissociation, ionization, and selective occupation of quantum states at low temperatures.  We extend these tables to the very low temperatures and pressures required for planets forming in the outer regions of disks (\App{EOStables}).
The EOS tables from \citet{saumon95} have often been used to model the interiors of giant planets (e.g., \citealt{pollack96}, \citealt{ikoma00}, \citealt{alibert05}, \citealt{hubickyj05}, \citealt{pn05}, \citealt{mordasini12}), as well as in complex stellar evolution simulations involving low temperatures (e.g., \citealt{paxton11}, \citealt{paxton13}, which use the code MESA). More recent EOS tables (\citealt{nettelmann08}, \citealt{nettelmann12}, \citealt{militzer13}) are based on ab initio molecular dynamics simulations and thus avoid some of the approximations of the \citet{saumon95} semi-analytical approach. However, though these newer tables are sufficient to model the internal structures of Jupiter and of close-in extrasolar planets,  they do not extend to the low temperatures and pressures required for wide-separation planets ($T \lesssim 500$ K, $P \lesssim 1$ GPa, e.g. \citealt{militzer13}). Moreover, the \citet{militzer13} EOS tables and the \citet{saumon95} tables are in good agreement for entropies, $S$, such that $\log_{10}(S) \gtrsim 8.75$ erg g$^{-1}$ K$^{-1}$.  All models presented here satisfy this constraint. 

\subsection{Opacity}
\label{sec:introopacity}

Opacities in protoplanetary disks are unlikely to be interstellar, and are lowered by grain growth and dust settling.  Numerous studies have demonstrated that atmospheric evolution depends strongly on opacity---lower opacities accelerate envelope growth and reduce $M_{\rm crit}$.  For example, the analytic model of \citet{stevenson82} showed that $M_{\rm crit} \propto \kappa^{3/4}$ for a fully radiative envelope with constant opacity $\kappa$. A series of more modern studies conclude that: Jupiter may have formed in 1 Myr with a core of $10 M_{\oplus}$ for an opacity arbitrarily reduced to 2\% of the interstellar value \citep{hubickyj05};  $M_{\rm crit}$ for Jupiter may be as low as $\sim$$1 M_{\oplus}$ for a grain free envelope \citep{hori10}; and the use of grain opacities that take into account coagulation and settling reduces the formation time for Jupiter by up to 80\%  \citep{movshovitz10}. A recent study by \citet{mordasini14} investigates the effect of opacity on the formation of populations of synthetic planets, and finds from comparisons with observations that opacities are likely to be much smaller than interstellar. Lastly, Paper I shows that reducing the opacity from interstellar to 1\% of interstellar reduces $M_{\rm crit}$ by a factor of $\sim$$2$.

Thus, to provide realistic estimates for $M_{\rm crit}$, we must employ realistic opacities.    For temperatures below dust sublimation, we employ opacity tables from \citet{dalessio01}, which are used to model observations of protoplanetary disks.  We present results for both a standard collisional cascade grain size distribution and for a shallower distribution, appropriate for coagulation (see Section \ref{sec:opacity} for details).  At high temperatures, where dust sublimates, we employ the analytic opacity law of \citet{bell94}.  For our purposes, this expression sufficiently represents the more detailed opacity tables of \citet{semenov03} with reduced computational complexity (see \App{radwindow}).  
We note that most previous works that study the quantitative effects of opacity reductions on $M_{\rm crit}$ are primarily focused on the formation of Jupiter at 5.2 AU.  In contrast,
our aim is to provide realistic estimates for $M_{\rm crit}$ for a larger parameter space, and to analyze the dependence of $M_{\rm crit}$ on semimajor axis.

%We use the EOS calculated by \citet{saumon95}, which captures non-ideal effects such as dissociation, ionization, and selective occupation of quantum states at low temperatures.  Opacities in protoplanetary disks are unlikely to be interstellar, and are lowered by grain growth and dust settling. We use the \citet{dalessio01} opacity tables, which are used to model observations of protoplanetary disks.

\subsection{Paper Plan}

After reviewing the quasi-static and cooling models derived in Paper I (Section \ref{sec2}), we add a non-ideal EOS.  Because the  adiabatic gradient, $\nabla_{\rm ad}$, is an important determinant of atmospheric structure, we first explain how dissociation and variable occupation of low-energy rotational states change $\nabla_{\rm ad}$ (Section \ref{deladtable}).  Section \ref{EOSeffects} presents the impact of these effects on atmosphere evolution.  We introduce realistic opacities in Section \ref{sec:opacity} and determine $M_{\rm crit}$ in Section \ref{critical}. Section \ref{acc}  compares our results to those obtained by studies employing high planetesimal accretion rates. We summarize our findings in Section \ref{conclusions}.

\section{Atmospheric Model Review}
\label{sec2}

%\textbf{Describe the assumptions of the model: disk, BCs, structure equations, cooling model etc., but with less detail than in paper I (and obviously refer to paper I for more details); for the BCs, emphasize that for a given core, the atmosphere profile and evolution are determined by the outer boundary conditions, i.e. Pout, Tout, Rout --- this will be relevant for section 3.2, i.e. outer boundary effects.}

We begin with a brief review of Paper I's model for the structure and evolution of a planetary atmosphere embedded in a protoplanetary disk. We summarize assumptions of the model and  properties of our assumed  disk in \S\ref{model} and list expressions for the atmosphere's structure and time evolution in \S\ref{struct}.  

\subsection{Assumptions and Disk Model}
\label{model}

We assume that the planet consists of a solid core of fixed mass and a two-layer atmosphere composed of an inner convective region and an outer radiative zone that matches smoothly onto the disk. The two regions are separated by the Schwarzschild criterion for convective instability (see \S\ref{struct}) at radius $r=R\cb$, known as the radiative-convective boundary (RCB). We assume that the luminosity is constant throughout the radiative region (see Section \S\ref{critical} for additional discussion). Note that a similar method is used by \citet{pn05} and \citet{mordasini12},  who find that it agrees with more complex models.

Our model applies when planetesimal accretion is minimal, so that the envelope's evolution is dominated by KH contraction (see \S\ref{acc} for a discussion of the physical conditions required for a core to be in this regime). The atmosphere is spherically symmetric, self-gravitating and in hydrostatic balance. We apply our model at semimajor axes $a\gtrsim5$ AU. Here the disk scale height is larger than the radius at which the planet matches onto the disk (see Paper I for further details), and hence spherical symmetry holds. The nebular gas is composed of a hydrogen-helium mixture, with hydrogen and helium mass fractions of 0.7 and 0.3, respectively. Because the envelope grows slowly, we calculate its evolution using a series of linked quasi-static equilibrium models.

%The time dependence of the atmosphere structure equations may be neglected or explicitly taken into account. Some previous studies of atmosphere accretion (e.g., \citealt{stevenson82}, \citealt{wuchterl93}, \citealt{rafikov06}) consider static envelopes, in which the luminosity is solely supplied by planetesimal accretion and fully radiated away by the atmosphere. In other studies, the time evolution is explicitly taken into account and full time dependent models are developed (e.g., \citealt{ikoma00}). We follow an intermediate approach and consider quasi-static evolution. Our model for the atmosphere growth time is described in section \ref{cooling}. 

The temperature and pressure at the outer boundary of the atmosphere are given by the nebular temperature and pressure. We use the minimum mass, passively irradiated disk model of  \citet{chiang10}. The surface density, mid-plane temperature and mid-plane pressure are

\begin{subeqnarray}
\label{eq:diskparam}
\Sigma\di&=&2200\, (a/\text{AU})^{-3/2}\,\, \text{g cm}^{-2} \slabel{eq:diska}\\
T\di &=& 120\, (a/\text{AU})^{-3/7} \,\,K \slabel{eq:diskb} \\
P\di&=&11\,  (a/\text{AU})^{-45/14} \,\, \text{dyn cm}^{-2} \slabel{eq:Pd} \, ,
\end{subeqnarray}
for a mean molecular weight $\mu=2.35$. 

\subsection{Structure Equations and Cooling Model}
\label{struct}

The structure of a static atmosphere is described by the standard equations of hydrostatic balance and thermal equilibrium:

\begin{subeqnarray}
\label{eq:struct}
\frac{d P}{d r}&=&-\frac{G m}{r^2}\rho \slabel{eq:structa} \\
\frac{d m}{d r}&=&4 \pi r^2 \rho\slabel{eq:structb} \\
\frac{d T}{d r}&=&\nabla \frac{T}{P}\frac{d P}{d r}\slabel{eq:structc} \\
\frac{d L}{d r}&=&4 \pi r^2 \rho (\epsilon + \epsilon_g)\slabel{eq:structd}, 
\end{subeqnarray}

\noindent where $r$ is the radial coordinate, $P$, $T$ and $\rho$ are the gas pressure, temperature, and density, respectively, $m$ is the mass enclosed by radius $r$, $L$  is the luminosity from the surface of radius $r$, and $G$ is the gravitational constant. The gas is heated at a rate $\epsilon_g \equiv -T dS/dt$ per unit mass due to gravitational contraction, where $S$ is the specific gas entropy, while  $\epsilon$ represents the rate at which internal heat is generated per unit mass. We do not take into account any internal energy sources and set $\epsilon=0$. The temperature gradient $\nabla \equiv d \ln T/d \ln P$ depends on whether energy is transported throughout the atmosphere by radiation or convection. In the case of radiative diffusion for an optically thick gas, the temperature gradient is

\begin{equation}
\label{eq:delrad}
\nabla = \delrad \equiv \frac{3 \kappa P}{64 \pi G m \sigma T^4} L,
\end{equation}

\noindent where $\sigma$ is the Stefan-Boltzmann constant and $\kappa$ is the dust opacity. In our models the atmosphere is optically thick throughout the outer boundary. Where energy is transported by convection, the temperature gradient is

\begin{equation}
\label{eq:delad}
\nabla = \delad \equiv \Big(\frac{d \ln T}{d \ln P}\Big)_{\mathrm{ad}},
\end{equation}
with $\delad$ the adiabatic temperature gradient. The convective and radiative layers of the envelope are separated by the Schwarzschild criterion (e.g., \citealt{thompson06}): the atmosphere is stable against convection when $\nabla < \delad$ and convectively unstable when $\nabla > \delad$. Since convective energy transport is highly efficient,  $\nabla \approx \delad$ in convecting regions. The temperature gradient is thus given by $\nabla=\mathrm{min}(\delad, \delrad)$. 

Equation set (\ref{eq:struct}) is supplemented by an equation of state (EOS) relating pressure, temperature and density, as well as an opacity law.  As summarized in Sections \ref{sec:introeos} and \ref{sec:introopacity}, this paper improves on the ideal gas polytropic EOS and interstellar medium dust opacity employed in Paper I.  We discuss the impact of our choices in Sections  \ref{deladtable}, \ref{EOSeffects}, and \ref{sec:opacity}.

We assume that the atmosphere forms around a solid core of fixed mass $M_{\rm c}$ with a radius $R_{\rm c}=(3 M_{\rm c}/4 \pi \rho_{\rm c})^{1/3}$, where $\rho_{\rm c}$ is the core density. We choose $\rho_{\rm c}=3.2$ g cm$^{-3}$ (e.g., \citealt{pap99}). Two radial scales determine the extent of the atmosphere: the Hill radius, $R_{\rm H} \equiv a [M_{\rm p} / (3 M_{\odot})]^{1/3}$, where the gravitational attraction of the planet and the tidal gravity due to the host star are equal, and the Bondi radius, $R_{\rm B} \equiv G M_{\rm p}/c_{\rm s}^2=G M_{\rm p} / (\mathcal{R} T\di)$, where the thermal energy of the nebular gas is approximately the gravitational energy of the planet. Here, $M_{\rm p}$ is the total planet mass, $c_{\rm s}$ is the isothermal sound speed, $\mathcal{R}=k_B/(\mu m_p)$ is the reduced gas constant, $k_B$ is the Boltzmann constant, and $m_p$ the proton mass. We define the planet mass as the mass enclosed inside the smaller of $R_{\rm B}$ or $R_{\rm H}$. For $R_{\rm B}<R_{\rm H}$, several studies assume that the atmosphere matches onto the disk at $R_{\rm B}$ (e.g., \citealt{ikoma00}, \citealt{pollack96}). In all cases, we choose the Hill radius as our outer boundary because the temperature and pressure at $R_{\rm H}$ are those of the disk,  $T(R_{\rm H})=T_{\rm d}$ and $P(R_{\rm H})=P_{\rm d}$ (see Paper I for more details). Outside $R_{\rm B}$, gas flows no longer circulate the planet, but rather belong to the disk. However, when $R_{\rm B}<R_{\rm H}$, the density structure between $R_{\rm B}$ and $R_{\rm H}$ remains spherical and is still well described by hydrostatic balance \citep{ormel13}.

Finally, we employ a cooling model developed in Paper I to determine the time evolution of the atmosphere between subsequent static models. A protoplanetary atmosphere embedded in a gas disk emits a total luminosity

\begin{equation}
\label{eq:coolingglobal}
L=L_c+\Gamma-\dot{E}+e_{\mathrm{acc}}\dot{M}-P_M \frac{\partial V_M}{\partial t}.
\end{equation}
Here, $L_{\rm c}$ is the luminosity from the solid core, which may include planetesimal accretion and radioactive decay, and $\Gamma$ is the rate of internal heat generation. We set $L\co=\Gamma=0$. The $\dot{E}$ term is the rate at which total energy (internal and gravitational) is lost.  Gas accretes at a rate  $\dot{M}(r)$ with a specific energy $e(r)=u(r) - G M(r) / r$, where $u(r)$ is the internal energy per unit mass and $M(r)$ is the mass enclosed by radius $r$. To evaluate Equation \ref{eq:coolingglobal}, we choose a boundary radius $r=R$, so that  $e_{\mathrm{acc}} = e(R)$.
%, and $e_{\mathrm{acc}}$ is the specific total energy brought in by gas accreting at rate $\dot{M}$: $e_{\mathrm{acc}}=u-G M/R$, where $u$ is the internal energy per unit mass. 
Finally, the last term in Equation \ref{eq:coolingglobal} represents the work done on a surface mass element.

%As a consequence of the equations above, both the atmosphere structure and the gas accretion rate are uniquely determined by the current atmosphere mass. As this mass accretion rate is slow compared to the time it takes to relax to this solution, we can make a quasi-static model of the atmosphere growth. 

We obtain an evolutionary series for the atmosphere by connecting sets of subsequent static atmospheres through the cooling Equation (\ref{eq:coolingglobal}). Details of our numerical procedure are described in Paper I. In contrast with Paper I, we evaluate Equation (\ref{eq:coolingglobal}) at $R_{\rm B}$ rather than $R\cb$. This ensures that our calculations are consistent,  because realistic opacities introduce inner radiative windows in the atmospheric structure and thus multiple RCBs (see Section \ref{sec:opacity}). 

%Paper I finds that atmosphere growth continuously accelerates after self-gravity becomes important, which happens at relatively low atmosphere masses, $M_{\rm atm} \lessim 0.1 M\co$. 

\section{Adiabatic Gradient for the Tabulated Equation of State}
\label{deladtable}

%\textbf{Explain the effects separately: dissociation vs. spin effects; show plots that explore these effects separately.}

%\subsection{Interpretation of Adiabatic Gradient Table}
%\label{deladinterp}

For our gas equation of state, we use the interpolated EOS tables of \citet{saumon95} for a helium mass fraction $Y=0.3$, and extend them to lower temperatures and pressures corresponding to the conditions in our fiducial disk. \App{EOStables} describes our extension procedure.
%useful parameter for understanding atmospheric structure, we choose to represent the EOS through $\delad$.

%devote this section to the impact on $\delad$ of the realistic EOS. 

%More details on our extension procedure and on the methodology of combining the separate tables for hydrogen and helium are presented in \App{EOStables}.

\begin{figure}[h]
\centering
\includegraphics[width=0.62\textwidth]{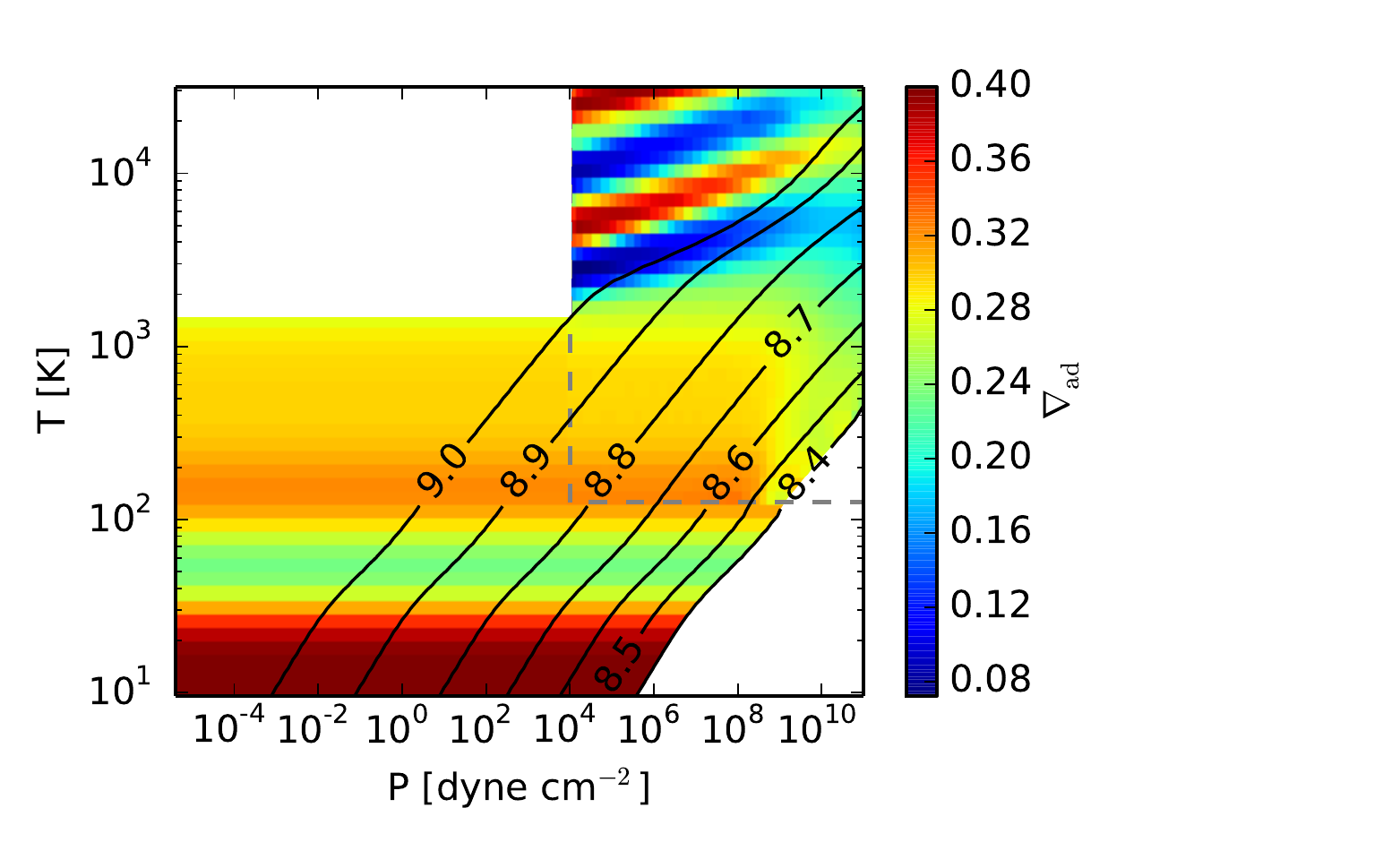}
%%\vspace{-0.5in}
\caption{Contour plot of the adiabatic gradient $\delad$ for a hydrogen-helium mixture in thermodynamic equilibrium as a function of gas temperature and pressure. The upper-right rectangle encloses the region described by the original \citet{saumon95} EOS tables, while the rest of the plot is our extension. The black curves represent constant entropy adiabats, with the labels $\log_{10}(S)$ [erg K$^{-1}$ g$^{-1}$]. The regions in which the EOS is either invalid or not computed are masked in white. Further details about the masked regions can be found in \App{hydrogen}.}
\label{fig:deladmap}
\end{figure}

 For ease of discussion, we choose to represent the EOS through the adiabatic gradient $\delad$ (defined in Equation \ref{eq:delad}). In contrast with an ideal gas polytrope, the EOS of a realistic gas cannot be fully specified by $\delad$. The \citet{saumon95} EOS tables are obtained using free energy minimization  (see, e.g., \citealt{graboske69}), which provides additional information about the relationship between the thermodynamic variables. Nevertheless, $\delad$ is a useful parameter for understanding atmospheric structure. For an ideal gas polytropic EOS, the adiabatic gradient is constant. Non-ideal effects such as dissociation or ionization produce temperature-dependent variations in $\delad$. Figure \ref{fig:deladmap} shows a contour plot of $\delad$  as a function of gas temperature and pressure. We distinguish three separate temperature regimes:

%, which was obtained by interpolating and extending the \citet{saumon95} EOS tables as described in \App{EOStables}.

%In this section we aim to explain how the variable adiabatic index of the hydrogen-helium mixture described by a real equation of state affects the atmosphere evolution when compared to an ideal gas with constant $\delad$. A contour plot of the adiabatic gradient as extrapolated from the \citet{saumon95} EOS tables is shown in Figure \ref{fig:deladmap}. We distinguish three separate regimes:

\begin{enumerate}
\item Intermediate temperature regime (300 K $\lesssim T \lesssim 2000$ K), where the hydrogen-helium mixture behaves like an ideal gas with a polytropic EOS.
\item High temperature regime ($T \gtrsim$ 2000 K), where dissociation of molecular hydrogen occurs, followed by ionization of atomic hydrogen for $T \gtrsim 10,000$ K.
\item Low temperature regime ($T \lesssim 300$ K), where the rotational states of the hydrogen molecule are not fully excited. % temperature is low enough for rotational motion to reduce.
\end{enumerate}

We note that helium behaves like an ideal monatomic gas with $\delad=2/5$ in our regime of interest.  Its presence in the atmosphere thus only causes a small, constant upward shift in the adiabatic gradient of the mixture.

%but not the white ones. We note, however, that the temperature and pressure ranges required by our models are fully covered by the colored regions, where our expressions are valid

%If follows that the expressions derived in \App{EOStables} cannot be used to extend the original tables in this temperature and pressure regime. 

%At high temperatures, hydrogen dissociates and ionizes, while at low temperatures the rotational states of the hydrogen molecule are only partially excited and it no longer behaves like an ideal diatomic gas.

%In what follows we explain the behavior of the adiabatic gradient in the three temperature regimes separately.

%\vspace{0.2in}

%\textbf{1. Intermediate T: Ideal Gas}

\subsection{Intermediate $T$: Ideal Gas}
\label{sec:intT}

For $300$ K $\lesssim T \lesssim 2000$ K, the hydrogen molecule is not energetic enough to dissociate and hydrogen behaves as an ideal diatomic gas with constant $\delad$ (Figure \ref{fig:deladmap}). The monatomic helium component increases the adiabatic index slightly, so that $\delad \approx 0.3$ rather than 2/7 for a pure diatomic gas.\footnote{Recall that for an ideal gas, $\delad \equiv \frac{\gamma-1}{\gamma}$, where the adiabatic index  $\gamma=7/5$ for a diatomic gas and 5/3 for a monatomic gas.}

%\vspace{0.2in}

%\textbf{2. High T: Dissociation and Ionization of Hydrogen.}

\subsection{High $T$: Dissociation and Ionization of Hydrogen}

 Hydrogen is molecular at low temperatures, dissociates at $T \gtrsim 2000-3000$ K, and ionizes at  $T \gtrsim 10,000$ K. In stellar and giant planet interiors there is little overlap between the two processes: hydrogen is almost entirely dissociated into atoms by the time ionization becomes important. 

As displayed in Figure \ref{fig:deladmap}, the adiabatic gradient decreases significantly in regions of partial dissociation and partial ionization.  This reduction occurs because dissociation and ionization act as energy sinks.  When internal energy is input into a partially dissociated gas, a portion is used to break down molecules, reducing the amount available to increase the temperature of the system. %The energy required to dissociate depends on the dissociation fraction, which can be determined from the Saha equation (see e.g., \citealt{kippenhahn90}). The dissociation fraction only depends on gas temperature and density, and hence only on the EOS. 

 %This behavior is different than that of a mixture of molecular and atomic hydrogen, for which $2/7<\delad<2/5$, or a mixture of protons and electrons, for which $\delad=2/5$. 

An expression for $\delad$ as a function of the dissociation fraction is presented in \App{deladdiss}. As expected,  $\delad$ is $2/7$ for pure molecular hydrogen and $2/5$ when hydrogen is fully dissociated, but decreases significantly during partial dissociation and is smallest when half of the gas is dissociated. Note that this behavior differs from a mixture of molecular and atomic hydrogen, for which $2/7<\delad<2/5$. %This behavior is shown in Figure \ref{fig:deladmap}. Ionization generates an analogous dip in $\delad$ at higher temperatures.

%The ionization of atomic hydrogen is also dictated by the Saha equation, with the dissociation energy replaced by ionization energy, hence the adiabatic gradient behaves analogously, consistent with Fig. \ref{fig:deladmap}.

%For a mixture of ideal gases, the total internal energy is given by the sum of the internal energies of the individual gases. 

%\vspace{0.2in}

%\textbf{3. Low T: Hydrogen Rotation and Spin Isomers}

\subsection{Low $T$: Hydrogen Rotation and Spin Isomers}

The (diatomic) hydrogen molecule has five degrees of freedom, three associated with translational motion and two associated with rotation. At $T$ well above the molecule's excitation temperature for rotation, $\Theta_r \approx 85$ K \citep{kittel}, its rotational states are fully excited and its EOS is that of an ideal diatomic gas (Section \ref{sec:intT}).   As $T \rightarrow 0$, rotation  ceases entirely and $\delad \approx 2/5$, the value for an ideal monatomic gas. %We note that at temperatures larger than $\gtrsim 6000$ K, vibrational motions also become important. While at $T\lesssim2000$ K, where our extension of the \citet{saumon95} EOS tables is valid, vibrational motion is negligible, we include these effects in our extension of the EOS for completion (see \App{EOStables} for details).
At  temperatures comparable to $\Theta_r$, the rotational states of $H_2$ are  are partially excited, and its EOS depends on the relative occupation of two isomeric forms, parahydrogen and orthohydrogen, distinguished by the spin symmetry of the molecule's two protons.  

Because protons are fermions, the Pauli exclusion principle implies that the total wave function of $H_2$ must be antisymmetric with respect to proton exchange.  Two components of the wave function may provide this asymmetry---proton spins and the molecule's rotational state. Parahydrogen has antiparallel proton spins and thus can only occupy symmetric rotational states with even angular quantum number $j$ \citep{farkas35}. In contrast, orthohydrogen has parallel proton spins, and can only occupy states with odd $j$. In thermal equilibrium at $T \rightarrow 0$ all hydrogen molecules are in the ground state with $j=0$, which corresponds to parahydrogen.  As temperature increases, parahydrogen starts converting into orthohydrogen.  Because the proton spin state is a triplet for orthohydrogen and a singlet for parahydrogen, this conversion plateaus at an ortho-para equilibrium ratio of 3:1 for $T \gtrsim 150$ K.
 
Spin conversion requires energy.  In thermal equilibrium, a portion of any internal energy input at $20 \lesssim T \lesssim 150$K is used to convert para- to orthohydrogen, reducing the amount available to increase $T$.  As a result, $\delad$ declines (Figure \ref{fig:deladmap}; see \App{deladspin} for further discussion). %At higher temperatures, the ortho-para 3:1 equilibrium ratio is reached; no further isomer conversion occurs, and thus $\delad$ remains relatively constant until dissociation temperatures are reached. 
 
\subsubsection{Thermodynamic equilibrium of spin isomers}

We conduct the majority of our calculations using the thermal equilibrium EOS illustrated in Figure \ref{fig:deladmap}.  
%In contrast, studies such as \citet{dangelo13} assume a fixed ortho-to-para ratio of 3:1 in their evolutionary calculations.  We include results for a fixed 3:1 ratio in Section \ref{critical} for reference (see \App{EOStables} and \ref{deladspin} for discussion of the EOS).
For reference, we also include results for a fixed 3:1 ortho-para ratio (used, e.g., by \citealt{dangelo13}).\footnote{When interconversion timescales are longer than the system evolution time, populations of the two isomers evolve independently, with a fixed abundance ratio.  Even at low temperature, formation of $H_2$ on grains produces an ortho-para ratio of 3:1 since the formation energy 4.48 eV, equipartitioned into vibrational, rotational, and translational energy, yields a rotation temperature of 9200K
%, which falls in the high temperature limit 
(e.g., \citealt{takahashi01}; c.f. \citealt{fukutani13}).  In the absence of thermal equilibrium, a fixed ratio of 3:1 is likely.}

Ortho- and parahydrogen remain in thermodynamic equilibrium if the isomers can interconvert on a timescale shorter than the atmosphere's KH contraction time.  In isolation, isomeric conversion requires a forbidden transition and has a timescale longer than the age of the universe \citep[e.g.,][]{pachucki08}.  Fast conversion requires a magnetic catalyst to aid the transition between the triplet and singlet spin states.

In astrophysical contexts, this catalyst is typically provided by collisions with ions such as $H^+$ or $H_3^+$ \citep[e.g.,][]{lique12,lique14}.  
%For example, collisions between $H_2$ and $H^+$ keep interstellar clouds in equilibrium \citep[e.g.,][]{dalgarno73}.  
At high densities, collisions with $H_2$ also contribute to conversion through interactions between the spin state and the magnetic dipole of the $H_2$ molecule \citep{huestis08}. 

%The low temperature EOS affects atmospheric evolution throughout the radiative envelope. It sets the scale height at the RCB, which determines the envelope's radiative diffusion time and hence the atmospheric luminosity (Equation \ref{eq:delrad} with $\nabla_{\rm rad} = \delad$).  The EOS above the RCB determines the scale length and hence the radial extent of the radiative region (see Section \ref{sec:roleofL}).  
For a core of mass $M_{\rm crit}$, the KH contraction time during the slowest phase of growth is approximately the disk lifetime, $t_{\rm disk} \sim 3\times 10^6$ years.  
The thermodynamic equilibrium EOS is thus appropriate if the ortho-para equilibration time, $t_{\rm equil}$ is smaller than $t_{\rm disk}$.
% in the outer atmosphere where $T \lesssim 150$ K.  
To determine whether this condition is met throughout the outer atmosphere (where $T \lesssim 150$ K), we check $t_{\rm equil}$ at the RCB and in the disk.
%To decide whether to use the thermodynamic equilibrium EOS, we therefore ask whether the ortho-para equilibration time $t_{\rm equil}$ is smaller than $t_{\rm disk}$ for $H_2$ number densities $n$ in a range spanning properties at the RCB and at the disk midplane, where the outer portion of the atmosphere matches smoothly.  

%In our models, RCB densities span the range $n  = 10^{13}$ to $5\times 10^{14}$ all times, cores 6--30 $M_\oplus$. Roughly dependent on the critical core mass.  Lower masses give higher densities. 1.5--5$\times 10^{14}$ during longest evolution   
For core masses 6--30$M_\oplus$, our models produce RCB densities in the range $n = $1.5--5$\times 10^{14}$ during the longest phase of atmospheric evolution.  The integrated column through the atmosphere at the RCB is $\sim$$10^{25}$cm$^{-2}$, larger than the $\sim$$10^{22}$--$10^{23}$ cm$^{-2}$ required to be optically thick to X-rays \citep{glassgold97}, so that ionizations in this interior portion of the atmosphere are attenuated. 
However, densities at the RCB are high enough that 
%even in the absence of an ion to act as a catalyst, 
$H_2$ collisions cause substantial isomeric conversion.
%can speed conversion to timescales shorter than the KH contraction time. 
Extrapolating from calculations of catalysis by $O_2$,  \citet{conrath84} estimate a rate coefficient for conversion of $k_{H_2} = 8\times 10^{-29} (T/125 {\rm K})^{1/2}$ cm$^3$ s$^{-1}$, 
%$k_{H_2} = (C/n) Z$, with $Z=4\times 10^{-19}$ and $(C/n) = 2\times 10^{-10} (T/125 {\rm K})^{1/2}$ cm$^3$ s$^{-1}$, 
marginally consistent with the production of non-equilibrium ortho-para ratios in solar system giants by upwellings and flows
 \citep[e.g.][]{fouchet03}. \citet{huestis08} estimates a faster rate coefficient given by $\log_{10} k_{H_2} [$cm$^3$ s$^{-1}] = 10^{-28}[1.56+12.2\exp(-173{\rm K}/T )]$, consistent with laboratory experiments in liquid and gaseous $H_2$ \citep{farkas35, milenko97}.  
 %The temperature in the radiative region only varies from the disk temperature by an order unity factor (see Section \ref{sec:roleofL}).  
 At $T = 50$K ($T$ in the radiative region only varies from the disk temperature by an order unity factor; see Section \ref{sec:roleofL}), the equilibration time is
\begin{equation}
t_{\rm equil} = (k_{H_2} n_{H_2})^{-1} = \text{1.5--6} \times 10^6 {\rm yr} \left(\frac{n}{10^{14} {\rm cm}^{-3}}\right)^{-1} \,\; ,
\end{equation}
which is comparable to or shorter than $t_{\rm disk}$.
%which, at stellocentric distance larger than FILL IN AU, is shorter than $t_{\rm disk}$ at the RCB for cores with mass $M_{\rm crit}$ during their slowest phase of atmospheric evolution. 

Where the outer atmosphere matches conditions in the disk, we turn to calculations for protoplanetary disks for guidance.  \citet{boley07}  estimate a minimum $t_{\rm equil} \sim 300$ yr.  Isomeric conversion primarily results from collisions between $H_2$ and ionic species, with $t_{\rm equil} \sim (k_{\rm ion} n_{\rm ion})^{-1} \sim 3\times 10^6$ yr $(n_{\rm ion}/10^{-4} {\rm cm}^{-3})^{-1}$, where $k_{\rm ion} \sim 10^{-10}$ cm$^3$ s$^{-1}$ \cite{walmsley04}.  Typical ion abundances in disks are small, and calculations require involved photochemical networks. 
%and are sensitive to assumptions about the properties of disk dust.  
Detailed work suggests that ion densities of $10^{-4}$ cm$^{-3}$ are achieved beyond 30 AU and, depending on the disk's dust complement, may be present throughout our region of interest \citep[e.g.,][]{glassgold97,bai09, turner10, perez11}. 
Inside $\sim$10AU, our atmospheric profiles are less sensitive to whether spin isomers reach thermodynamic equilibrium since disk temperatures exceed the peak conversion temperature of $\sim$50K.

We conclude that during the longest phase of atmospheric growth, to which $M_{\rm crit}$ is most sensitive, the equilibrium EOS is appropriate for the majority of and possibly all of our parameter space.

For our calculations, the difference between thermal equilibrium and a fixed ortho-para ratio of 3:1 is more prominent at larger stellocentric distances, where disk temperatures are lower. We find  that a fixed ortho-para ratio may increase the atmospheric evolutionary time by up to a factor of $\sim$$3$ (corresponding to 100 AU in our fiducial disk; see \App{deladspin}, Figure \ref{fig:Lt_31}), and hence $M_{\rm crit}$ by up to a factor of 2. This effect is much smaller in the inner parts of the disk --- the atmospheric growth time only increases by $\sim$$25$\% at 10 AU.

 %At equilibrium, the relative abundance of the ortho- and para- states is given by the ratio of their partition functions, described in \App{EOStables}.
%During the para-to-ortho conversion, part of the internal energy of the hydrogen molecule is used 

\section{Role of the Equation of State}
\label{EOSeffects}

Variations in the EOS, and hence $\delad$, due to partial dissociation and fractional occupation of $\rm{H}_2$ rotational states affect atmospheric evolution by yielding: (1) a lower envelope luminosity $L$, and (2) a larger amount of radiated energy per unit of accreted mass $-dE/dM$, when compared to the polytropic EOS considered in Paper I. As a result, the rate of change in atmospheric mass,

\begin{equation}
\label{eq:dMdt}
\dot{M} = -\frac{L}{dE/dM},
\end{equation}
(cf. Equation \ref{eq:coolingglobal} and ignoring surface terms, which only become significant in the late stages of atmospheric evolution) is lower than in the polytropic case. Slower gas accretion increases the growth time of the atmosphere. Since envelope growth is faster for larger cores, we calculate a larger $M_{\rm crit}$. % --- the minimum core mass required to accumulate a massive envelope during a fixed disk lifetime.%, i.e. $M_{\rm crit}$, increases when realistic EOS effects are included.

Modifications in the EOS affect $L$ and $-dE/dM$ because both of these quantities depend on the global structure of the envelope. From \Eq{eq:delrad} applied at the RCB where $\delad=\delrad$, and for a fixed atmospheric mass, the luminosity emerging from the convective interior scales as $L \propto T\cb^4/P\cb$. We have shown in Paper I that the outer radiative layer of the atmosphere is nearly isothermal, and that $T\cb$ is only a factor unity correction from the disk temperature, while the pressure in the radiative region increases exponentially with depth. It follows that $L$ primarily scales as $1/P\cb$. This pressure depth at the RCB depends on both the interior structure of the atmosphere at high temperatures, and on the matching to the nebula through the radiative region at low temperatures. Similarly, the EOS affects the distribution of total energy $E(r)$ throughout the envelope, and thus $-dE/dM$.

Since both dissociation deep in the convective interior and variable occupation of rotation states in the outer envelope affect atmospheric structure and evolution, it is helpful to separate these effects and study them independently.

\begin{figure}[h]
\centering
\includegraphics[width=0.5\textwidth]{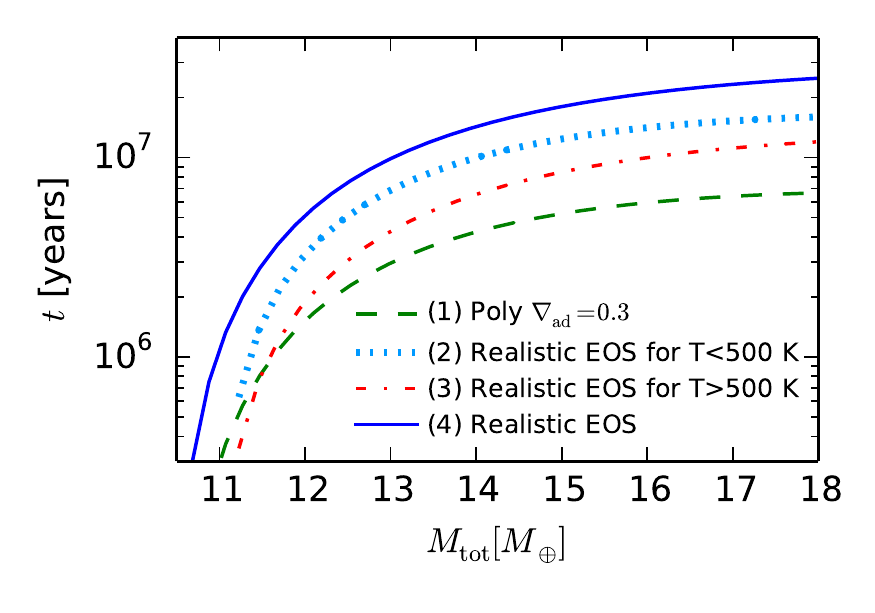}
%%\vspace{-0.5in}
\caption{Elapsed time to grow a planet of total mass (core + atmosphere) for a variety of EOS combinations (see text), for a planet forming at 5 AU and with a fixed core mass $M_{\rm c}=10 M_{\oplus}$. Both hydrogen dissociation at high temperatures deep in the atmosphere and fractional occupation of $\rm H_2$ rotational states at low temperatures in the outer envelope result in slower gas accretion when compared to an ideal gas polytrope.}
\label{fig:tplotall}
\end{figure}

Figure \ref{fig:tplotall} shows the time evolution of atmospheres forming at 5 AU around cores of mass $M\co=10 M_{\oplus}$ and described by various equations of state, as follows:
%{\bf I would make this list numbered and omit the reference to the curve styles, which is clear in figures.}
\begin{enumerate}
\item Ideal gas polytrope with $\delad=0.3$. % (hereafter EOS 1).
\item Ideal gas polytrope with $\delad=0.3$ for $T>500$ K and realistic EOS for $T<500$ K, which includes the effects of fractional occupation of $\rm H_2$ rotational states.
\item Ideal gas polytrope with $\delad=0.3$ for $T<500$ and realistic EOS for $T>500$ K , which includes the effects of hydrogen dissociation.
\item Realistic EOS at all $T$. 
%({\bf Here and elsewhere, I don't like ``fully" realistic, as it implies a perfect EOS.  I would just say realistic at all $T$.})
\end{enumerate}
We choose $T=500$ K as the cutoff temperature because the hydrogen-helium mixture behaves like an ideal gas, with $\delad=0.3$, in this temperature regime (see Figure \ref{fig:deladmap}). 
%{\bf (I find the references to curve numbers distracting in general.  I would try to do less of this and focus on describing the physical effects.  The reader still needs to look at the figure, and there the curves are clearly labelled.)} \sout{As such}, 
%By separately comparing curves (1) and (2), and (1) and (3)  while the difference between curves (1) and (3) accounts for hydrogen dissociation. 
Both EOS 2 and EOS 3 yield slower gas accretion when compared to EOS 1. Noting the logarithmic scale in Figure \ref{fig:tplotall}, we see that the combined effect of fractional occupation of rotational states and dissociation is significantly greater than either individually. We explore these contributions in Sections  \ref{sec:dEdM} and \ref{sec:roleofL}.

% --- due to dissociation for the high-$T$ EOS, and due to fractional occupation of $\rm H_2$ rotational states for the low-$T$ EOS.
%Both dissociation and fractional occupation of $\rm H_2$ rotational states result in slower cooling for both the high-$T$ and low-$T$ realistic EOSs. 
%decrease the atmospheric accretion rate $\dot{M}$ and thus
%This follows from Equation (\ref{eq:dMdt}), with rotation decreasing $L$ and dissociation increasing $-dE/dM$, as we show further. 
%consistent with Equation \ref{eq:dMdt}, since dissociation increases $-dE/dM$ while molecular rotation decreases $L$.
  
 %From Equation \ref{eq:dMdt}, the atmospheric growth time is dependent on 

%The growth time is dependent on both the total energy that must be released (i.e., the rate at which energy is released) and on the luminosity of the atmosphere. {\bf `Rate of energy release' and luminosity are the same thing.  You may be calling $dE/dM$ a `rate' which is confusing.  Again, I think if you introduce and physically explain $\dot{M} = -L/(dE/dM)$ at the outset, then the resulting explanations can more compactly and clearly refer to the terms that matter.}  We explore the \sout{relative} influence of these two factors separately.  {\bf Haven't you already started exploring these factors separately?  Why say this again? Also perhaps worth a note (if not an explanation) that the combined effect is significantly greater than either individually (again noting log scale).}

\begin{figure*}[tb]
\centering
\includegraphics[width=\textwidth]{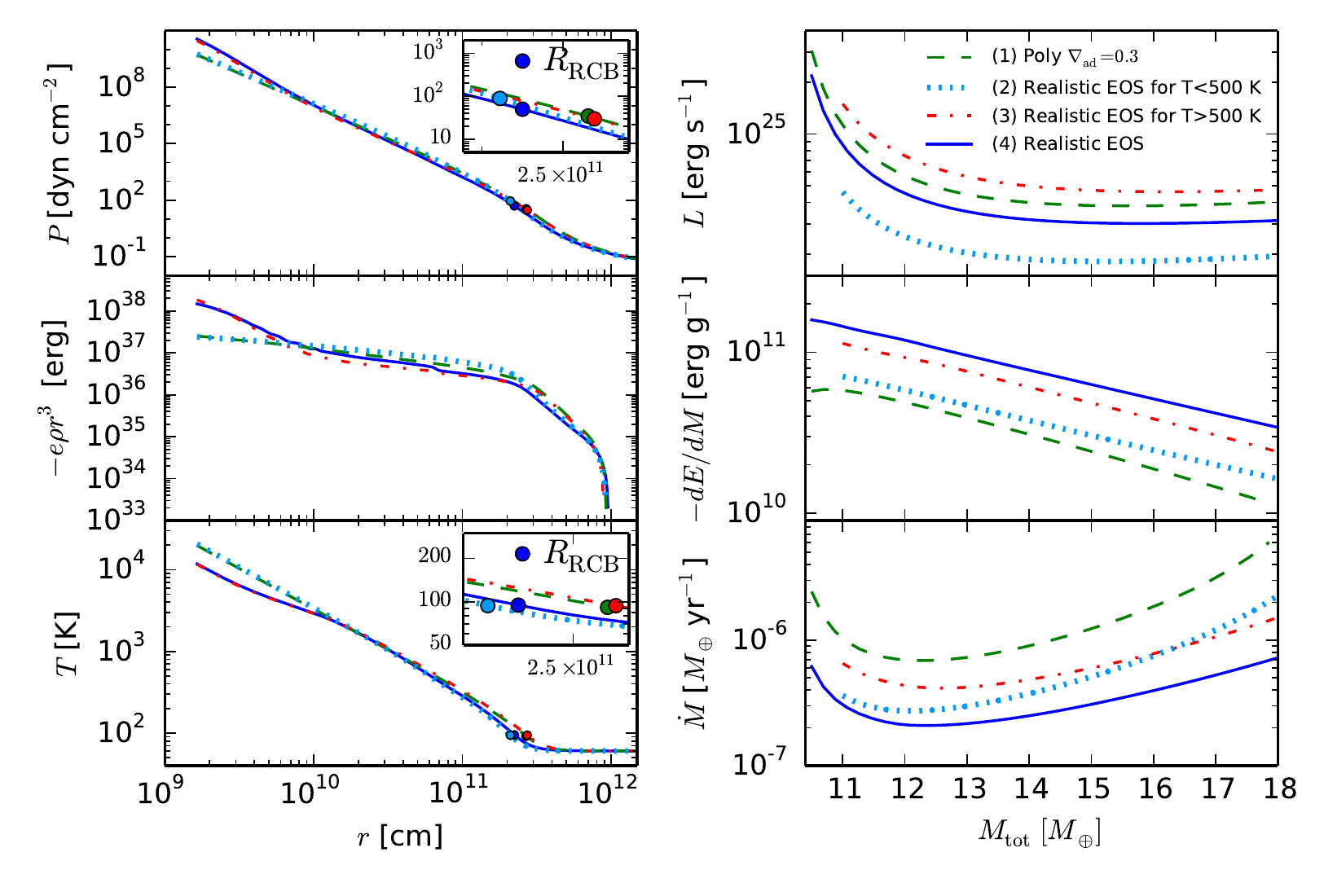}
%%\vspace{-0.5in}
\caption{
%{\bf (Left and right panels could be different figs.  The left panel could benefit from showing $P$ or $\rho$.  Also $E_{\rm tot}$ is not yet well defined; the local quantity $e \rho r^3$ might be more instructive.  A greater range of $E$ would show what the polytrope is doing better.  The right panel could benefit from adding $\dot{M}$.  We call $E_{\rm tot} \rightarrow E$ elsewhere, so should be consistent.  Why don't x-axes line up in right panel, are masses different?  What is the discontinuity in $dE/dM$ for the polytrope around 14 $M_\oplus$?  (A kink would be course resolution, but a discontinuity is harder to understand.)  I don't think you need an abs.\ val. around $-dE/dM$.)} 
We explore atmosphere growth around a core of $M\co=10 M_{\oplus}$ forming at 5 AU in our fiducial disk, for the EOS choices in Figure \ref{fig:tplotall}. Left panels show radial profiles of pressure (upper), $-e \rho r^3$ (middle) and temperature (lower), for a total mass (core + atmosphere) of $12 M_{\oplus}$. Right panels display evolution with total mass of $L$ (upper), $-dE/dM$ (middle) and $\dot{M}$ (lower). Upper-left and upper-right: The variable occupation of $\rm H_2$ rotational states in the outer atmosphere results in a deeper RCB and a lower luminosity for the realistic EOS. Middle-left and middle-right: the (negative) total energy of the atmosphere is more concentrated at the bottom of the envelope when compared to an ideal gas polytrope due to $H_2$ dissociation. This increases the amount of energy per unit mass, $-dE/dM$, that needs to be radiated away to accrete the next parcel of gas. Lower-left: $H_2$ dissociation in the inner atmosphere decreases the temperature near the core. Lower-right: Dissociation and fractional occupation of rotational states reduce $\dot{M}$ and thus slow down atmospheric growth.}
\label{fig:all_plot}
\end{figure*}

%In Figure \ref{fig:all_plot}, middle-right and upper-right panels, we show an example evolution with planet mass of $-dE/dM$ and $L$, respectively, for the EOSs described above. The realistic EOS increases $-dE/dM$ by a factor of $\sim$$3$ and decreases $L$ by a factor of $\sim$$1.5$, when compared to the ideal gas polytrope. 
%Atmospheric growth is therefore primarily slowed by changes in the energy profile $E(r)$, with the luminosity $L$ having a secondary effect. 
%In what follows we explore each contribution separately. 

Throughout Section \ref{EOSeffects}, we use a power-law opacity given by
\begin{equation}
\label{eq:opacitylaw}
\kappa=2 F_{\kappa} \Big(\frac{T}{T_{\rm ref}}\Big)^{\beta},
\end{equation}  
\noindent with $\beta =2$, $F_{\kappa} = 1$, and $T_{\rm{ref}}=100$K, appropriate for ice grains in the interstellar medium (ISM) \citep{bell94}.  We improve our treatment of opacity in Section \ref{sec:opacity}.

\subsection{Role of $-dE/dM$}\label{sec:dEdM}

Variations in $\delad$ due to dissociation increase $-dE/dM$ (Figure \ref{fig:all_plot}, middle-right). 
The atmosphere's total (negative) energy at scale $r$, $-e \rho r^3$ 
%Figure \ref{fig:all_plot}, middle-left panel, shows a radial profile of the atmosphere's total (negative) energy at scale $r$, $-e \rho r^3$, where $e$ is the total specific energy.   %the cumulative total energy (internal and gravitational) profile as a function of the radial coordinate, \sout{for the same example planet}. {\bf Two choices: (1) define this cumulative energy more clearly and give it a different symbol than $E_{\rm tot}$ that's already used, maybe $E(r)$, or (2) plot the local (noncumulative) quantity $e \rho r^3$, which should peak in the interesting places.} 
%The bulk of the energy 
is concentrated in the atmospheric interior for all EOSs (Figure \ref{fig:all_plot}, middle-left). However, $-e \rho r^3$ for EOS 3 and EOS 4 is much larger in magnitude near the core when compared to the others. This is due to the fact that $\delad$ decreases in dissociation regions (Figure \ref{fig:deladmap}). Dissociation adds particles to the gas and hence increases its entropy. In order for entropy to stay constant with radius in the convective interior, the temperature must drop, lowering $\delad$ and $|dT/dr|$ and resulting in lower temperatures near the core (cf. Figure \ref{fig:all_plot}, bottom-left). Because  the specific internal energy $u \propto T$, dissociation decreases the total energy deep in the interior.  Note that this loss of thermal energy due to dissociation can be large enough to trigger dynamical instabilities and eventual collapse in higher mass objects such as protostars \citep{larson69} or during the runaway growth of giant planets \citep{bodenheimer80}.

The reduced total energy of an atmosphere with a realistic EOS can also be explained qualitatively using $\delad$.  The density profile in an adiabatic, non-self-gravitating atmosphere composed of an ideal gas scales as $\rho(r) \propto r^{-1/\delad+1}$, and the total specific energy is $e(r) \propto 1/r$ (see Paper I). Thus $- e \rho r^3 \sim r^{-1/\delad+3}$.
 The total energy is concentrated near the core if $\delad<1/3$, and at the outer boundary otherwise. 
 %In our regime of interest, $\delad<1/3$ for all EOS choices, so the total energy  is concentrated near the core (cf. Figure \ref{fig:all_plot}, middle-left).  
Dissociation reduces $\delad$, and for smaller $\delad$, the total energy is more tightly packed in the envelope's interior.

%Dissociation reduces $\delad$, causing $-e \rho r^3$ to rise more sharply with radius. Smaller $\delad$ 
%More importantly, the dependence of $-e \rho r^3$ on $\delad$ also implies that adiabats with lower $\delad$ have their energy more tightly packed towards the interior of the envelope, which is the case during dissociation. 

An atmosphere with more negative total energy (EOS 3 and 4) requires a larger amount of energy to be radiated away to bind the next batch of gas. This implies a larger $-dE/dM$ during atmospheric evolution (Figure \ref{fig:all_plot}, middle-right), and thus a lower $\dot{M}$ (Figure \ref{fig:all_plot}, lower-right).

\begin{figure*}[tb]
\centering
\includegraphics[width=\textwidth]{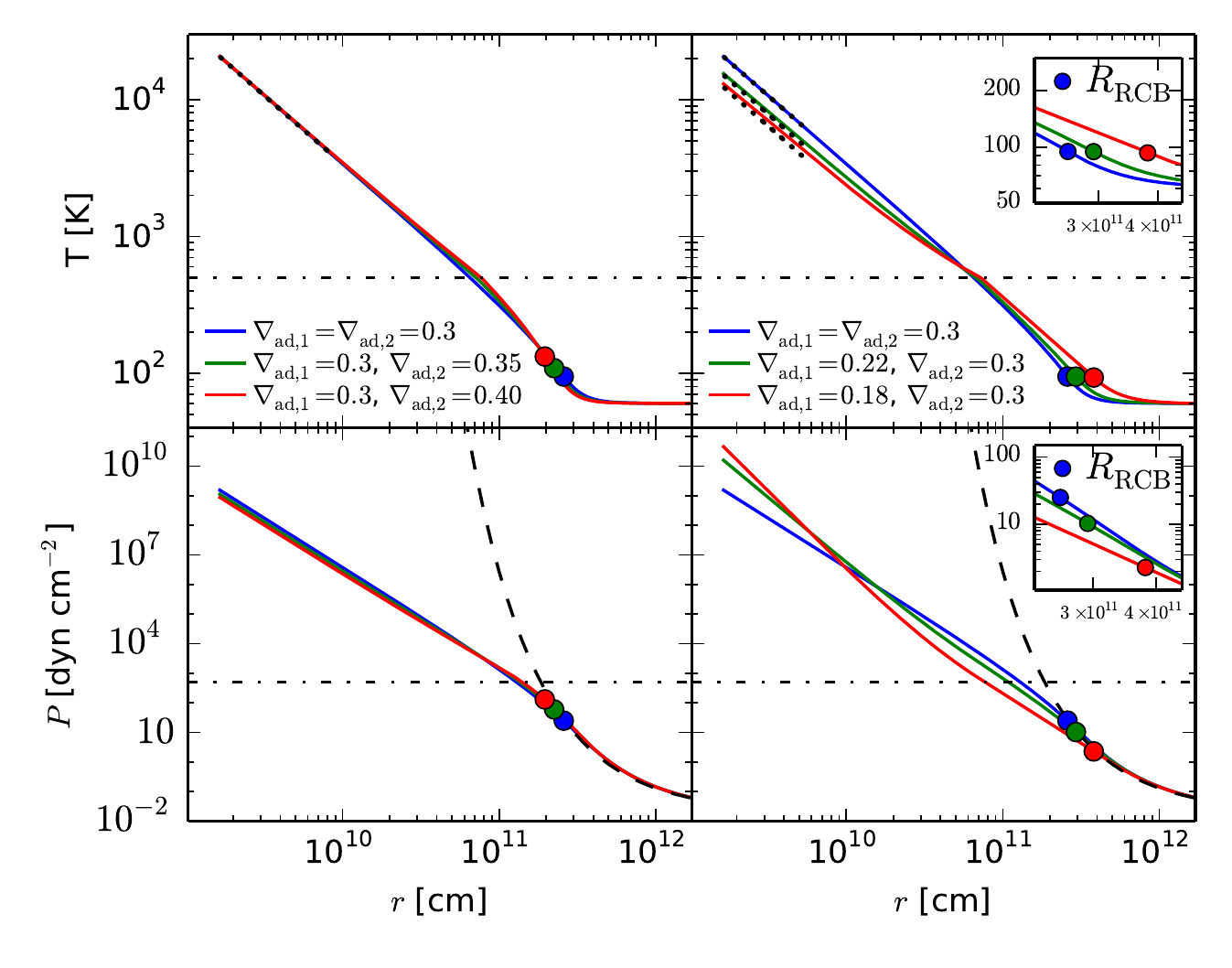}
%%\vspace{-0.5in}
\caption{As an illustrative numerical experiment, we explore atmosphere growth around a core with $M\co=10 M_{\oplus}$ forming at 5 AU, and with total mass (core + atmosphere) 12 $M_{\oplus}$, for the EOS choice of \Eq{eq:deladvar} and various $\nabla_{\rm ad,1}$ and $\nabla_{\rm ad,2}$. In each panel, the circles mark the RCB location. The horizontal dash-dotted lines show the temperature $T_{\rm switch}=500$ K at which the adiabatic gradient changes. We show radial profiles of temperature (upper panels) and pressure (lower panels). Upper-left: fixed $\nabla_{\rm ad,1}$, varying  $\nabla_{\rm ad,2}$. A larger $\nabla_{\rm ad,2}$ yields a deeper radiative region. Upper-right: varying $\nabla_{\rm ad,1}$, fixed $\nabla_{\rm ad,2}$. A lower $\nabla_{\rm ad,1}$ yields a more shallow radiative region. In both upper panels, the analytic profile of Equation (\ref{eq:Tdeep}) is over-plotted for comparison (\textit{dotted line}) in regions of agreement. Lower-left: fixed $\nabla_{\rm ad,1}$, varying  $\nabla_{\rm ad,2}$. Lower-right: varying $\nabla_{\rm ad,1}$, fixed $\nabla_{\rm ad,2}$. In both lower panels, the dashed black line is the analytic prediction for the pressure profile in the radiative zone, which agrees with the numerical results.}
\label{fig:varying_delad}
\end{figure*}

\subsection{Role of $L$}\label{sec:roleofL}

%The radiative and convective profiles must match at the RCB, constrained by the known total mass of the atmosphere

%The EOS variations in the cold outer envelope affect $\dot{M}$ by decreasing the atmosphere's luminosity compared to an ideal gas polytrope (Figure \ref{fig:all_plot}, upper-right panel, for the same parameters as Figure \ref{fig:tplotall}). From Figures \ref{fig:deladmap} and \ref{fig:all_plot}, bottom-left panel, we see that the adiabatic gradient of the low-$T$ realistic EOS increases with radius (i.e., with decreasing temperature) throughout the convective zone when $T<500$ K, then becomes lower near the RCB. This results in a larger adiabatic gradient, on average, for the low-$T$ realistic EOS in the outer parts of the convective region of the atmosphere. 

%The variations in $\delad$ at low temperatures 

The increase in $\delad$ at low temperatures, where $H_2$ rotational states are not fully excited (Figure \ref{fig:all_plot}, upper-right) decreases the atmosphere's luminosity.
%compared to an ideal gas polytrope 
This result may be understood as consequence of matching between the atmosphere's interior convective and exterior radiative profiles.  The profiles must match at the RCB, constrained by the known total mass.  This match determines the pressure at the RCB, $P\cb \propto \exp(\RB/R\cb)$. Higher $P\cb$ implies lower luminosity (Equation \ref{eq:delrad}).

The adiabatic gradient of EOS 2 is larger, on average, than for EOS 1 in the outer part of the convective region.\footnote {$\delad$ for EOS 2 increases with radius (i.e., with decreasing temperature) throughout the convective zone, then becomes lower near the RCB.} To understand how a variable, but overall larger $\delad$ in the outer atmosphere affects evolution, we study the simplified problem of atmospheres composed of an ideal gas with adiabatic gradient 

\begin{equation}
\label{eq:deladvar}
\delad = \left\{
\begin{array}{l l}
\nabla_{\rm ad, 1}, & \quad T > 500 \text{ K} \\
\nabla_{\rm ad, 2}, & \quad T < 500 \text{ K}
\end{array} 
\right.
\end{equation}   
where $\nabla_{\rm ad, 1}$ and $\nabla_{\rm ad,2}$ are constant (Figure \ref{fig:varying_delad}).
%We explore the independent effects of varying $\nabla_{\rm ad,1}$ and $\nabla_{\rm ad,2}$ by alternately fixing one and varying the other   (Figure \ref{fig:varying_delad}). 
%The adiabatic gradient $\nabla_{\rm ad,1}$ dictates the temperature profile deep in the convective interior, while $\nabla_{\rm ad,2}$ sets the RCB temperature $T\cb$. 
%Given these inputs, the RCB location and its pressure $P\cb \propto \exp(\RB/R\cb)$ (see Figure \ref{fig:varying_delad}, lower panels) are determined by matching the interior and exterior profiles. Higher $P\cb$ implies lower luminosity (Equation \ref{eq:delrad}).

%the location of the RCB, is set by the matching between the convective interior and the outer radiative zone. As $P(r) \propto \exp(\RB/r)$ in the radiative region, 

%As $P(r) \propto \exp(\RB/r)$ in the outer radiative region, 

%Figure \ref{fig:varying_delad}, lower-left panel, shows the radial pressure profile for $\nabla_{\rm ad,1}=0.3$ and various $\nabla_{\rm ad,2}$, for the same parameters as Figures \ref{fig:tplotall} and \ref{fig:all_plot}, which are in agreement with the analytic expectation of an exponential profile. 

%The numerical solutions reproduce the analytic expectation of a nearly exponential pressure profile in the radiative region, $P(r) \propto \exp(\RB/r)$. Thus $L \propto 1/P\cb \propto 1/\exp(\RB/R\cb)$ is directly correlated to the physical depth of the RCB. 

%The location of the RCB is set by the matching between the convective interior and the outer radiative zone. 

The adiabatic gradient $\nabla_{\rm ad,1}$ dictates the temperature profile deep in the convective interior.  Deep in the atmosphere, where $r \ll R\cb$, virial equilibrium yields a temperature profile 
\begin{equation}
\label{eq:Tdeep}
T(r) \approx \frac{\nabla_{\rm ad,1} G M\co}{\mathcal{R} r}.
\end{equation}   
Even though $\nabla_{\rm ad,1}$ is an order unity coefficient, the temperature scaling with $\nabla_{\rm ad,1}$ in \Eq{eq:Tdeep} is exact (Figure \ref{fig:varying_delad}, upper-right panel). Because the radial temperature profile in the atmospheric interior is set by $\nabla_{\rm ad,1}$, the radius $R_{\rm switch}$ at which the adiabatic gradient changes shows little variation with $\nabla_{\rm ad,2}$. %We can therefore match the fixed interior temperature profile to the exterior in order to determine $R\cb$ as a function of $\nabla_{\rm ad,2}$. Note, however, that \Eq{eq:Tdeep} is only valid to order of magnitude at $R_{\rm switch}$.

%n contrast with the deep interior, however, we cannot rigorously apply \Eq{eq:Tdeep} at $R_{\rm switch}$, although the approximation is still valid to order unity.

%We can use 
The RCB temperature, $T\cb$, is an order unity correction from the disk temperature that depends on the adiabatic gradient in the outer envelope, $\nabla_{\rm ad,2}$.  We have shown in Paper I that $T\cb \simeq T\di (1- 2 \nabla_{\rm ad, 2})^{-1/2}$ (for $\beta=2$ in Equation \ref{eq:opacitylaw}). This approximation agrees with the numerically calculated $T\cb$ shown in Figure \ref{fig:varying_delad} to within 5\%. Note that $T\cb$ modestly increases with $\nabla_{\rm ad,2}$ (upper-left panel), and is constant for fixed $\nabla_{\rm ad,2}$  regardless of $\nabla_{\rm ad,1}$ (upper-right panel).

% modestly increases with $\nabla_{\rm ad,2}$, consistent with the upper-left panel of Figure \ref{fig:varying_delad}. Moreover, a constant $\nabla_{\rm ad,2}$ in the exterior yields the same value for $T\cb$, independent of $\nabla_{\rm ad,1}$ (Figure \ref{fig:varying_delad}, upper-right panel), again consistent with the analytic prediction. Quantitatively, we have found that the analytic approximation for $T\cb$ agrees with the numerical result to within 5\%.
When these temperature conditions are matched for a fixed atmosphere mass, Figure \ref{fig:varying_delad} shows that a larger adiabatic gradient in the outer atmosphere results in a deeper radiative region. We can understand this effect by assuming that $R_{\rm switch}$ is fixed. A steeper adiabat (i.e., with a larger adiabatic gradient) that starts at a given depth (here $R_{\rm switch}$) will match the fixed RCB temperature at a smaller radius. A quantitative scaling of $R\cb$ is not possible because Equation (\ref{eq:Tdeep})  is only valid to order of magnitude at $R_{\rm switch}$ and relatively small variations in the physical depth of the RCB can impact the RCB pressure significantly. It follows that $L \propto 1/P\cb$ decreases with increasing outer $\nabla_{\rm ad}$; thus the overall larger adiabatic gradient due to the variable occupation of rotation states at low temperatures decreases $L$ and slows down growth.

\section{Impact of Opacity on Atmosphere Evolution}
\label{sec:opacity}

Our calculations so far have assumed that the dust opacity in the radiative region of the atmosphere is given by the standard ISM opacity. However, our scenario of low planetesimal accretion is likely to favor lower dust opacities, due to grain growth and dust settling. Grain growth, in particular, lowers the absolute value of the opacity, and may change the particle size distribution when compared to the standard ISM size distribution (e.g., \citealt{pollack85}). Enhanced metallicity due to planetesimal accretion, by construction not present during our atmosphere's growth, cannot make up for this reduction.

%Observations of dust in protoplanetary disks (e.g., \citealt{beckwith90}, \citealt{beckwith91}, \citealt{perez12}) have shown evidence for grain growth and that the particle size distribution is not interstellar. 

Although grain growth and evidence for a non-ISM size distribution have been observed in protoplanetary disks (e.g., \citealt{beckwith90}, \citealt{beckwith91}, \citealt{perez12}), the size distribution of dust particles has not been tightly constrained. Typically, the differential grain size distribution is assumed to be a power-law: 

\begin{equation}
\label{eq:graindistr}
\frac{dN}{ds} \propto s^{-p},
\end{equation}
where $dN$ is the number of particles with sizes between $s$ and $s + ds$ and $p=3.5$ (a standard \citealt{dohnanyi69} collisional cascade; appropriate for the ISM) or $p=2.5$ (an approximation for coagulation). In this work we use the  \citet{dalessio01} frequency-dependent opacity tables to obtain the temperature-dependent Rosseland mean opacity $\kappa$. We take as a fiducial case a maximum particle size $s_{\rm max}=1$ cm and a grain size distribution given by Equation (\ref{eq:graindistr}) with $p=3.5$. Other choices for the power-law coefficient $p$ are discussed later in this section. 
Though we choose our opacities to reflect ambient conditions in the disk, we note that 
%even if the opacity were close to that of the ISM in the disk, 
grains can also grow within an accreting atmosphere, decreasing its opacity at depth \citep{movshovitz10,mordasini14b,ormel14}.

The \citet{dalessio01} opacities are only relevant at temperatures that are sufficiently low for dust grains to remain solid ($T \lesssim 1000$ K). At higher temperatures, we use the \citet{bell94} analytic opacity laws, ensuring smooth transition from the grain growth opacities, as illustrated in Figure \ref{fig:opacity}. 

The sharp drop in opacity ($\kappa \sim T^{-24}$, see Figure \ref{fig:delvsr}) due to dust sublimation lowers the radiative temperature gradient significantly (see Equation \ref{eq:delrad}), and may thus generate radiative layers within the inner region of the atmosphere (see \App{radwindow}, Figure \ref{fig:delvsr}). Additionally, the weak temperature dependence of grain growth opacities may cause larger outer radiative zones than when ISM opacities are used. This could pose two challenges for our model:

\begin{enumerate}

 \item The additional luminosity, $\Delta L$, generated in the outer radiative layer of the envelope may not satisfy $\Delta L \ll L$, where $L$ is the assumed fixed atmospheric luminosity. For $p=3.5$ in Equation (\ref{eq:graindistr}), we have checked that $\Delta L \ll L$ in all the cases presented in this study. For $p=2.5$, however, this approximation breaks down at low core masses, as we show in \S\ref{critical}.

 %In practice, however, the inner radiative windows are either very narrow compared to the height of the convective regions (Figure \ref{fig:delvsr}, middle panel), or  have $\delad \approx \delrad$ throughout, which makes the distinction between convective and radiative layers less  pronounced (Figure \ref{fig:delvsr}, bottom panel). 

%This results in a negligible extra luminosity generated in the radiative windows. However, the radiative windows may become non-negligible for other opacity choices and sufficiently low core masses, as we show later in this section.

\item As little as half of the atmosphere's luminosity is generated in the innermost convective layer when radiative windows exist.   We must therefore check that our assumption of constant $L$ does not substantially change the structure of the atmosphere in the region of the radiative windows. Fortunately, these radiative windows are either very narrow compared to the height of the convective regions (Figure \ref{fig:delvsr}, middle panel), or  have $\delad \approx \delrad$ throughout, which makes the distinction between convective and radiative layers less  pronounced (Figure \ref{fig:delvsr}, bottom panel). This implies that the entropy drop across the radiative windows is small. We investigate the luminosity structure in greater detail in \App{radwindow} and conclude that our model remains a reasonably good approximation even in the presence of radiative windows.
%\textit{We have verified using an extreme luminosity profile that our results reasonably approximate the atmosphere's structure interior to the outermost radiative layer (see \App{radwindow} for additional details)}.

%By design, the luminosity in the radiative windows, $L_{\rm radw}$, must satisfy $L_{\rm radw}=L$, where $L$ is the assumed fixed luminosity at the top of the atmosphere. Non-negligible extra luminosity generated in the outer convective layers would, however, yield $L_{\rm radw}<L$, in conflict with our assumptions, and could change atmospheric structure. A simple way to check this \textit{a posteriori} is by rewriting the local energy equation (\ref{eq:structd}) as $\partial L/\partial m=-T \partial S/\partial t$, and integrating it throughout the atmosphere assuming that $\partial S/\partial t$ is fixed (e.g., \citealt{arras06}). If the value of this integral, $I$\footnote{This integral does not have units of luminosity, as we drop the constant $\partial S/\partial t$ term.}, throughout the innermost convective region is significantly larger than its value throughout the rest of the envelope, $\Delta I$, then our assumptions hold. For all the models for which $\Delta L \ll L$ holds (see paragraph above), we have found $\Delta I / I \lesssim 30\%$, and typically $\lesssim 10\%$ (see \App{radwindow} for additional details).

\end{enumerate}

We thus find that our atmospheric and cooling model is valid in our regions of interest, with some exceptions discussed in \S\ref{critical}. %Due to the variable number and position of radiative windows, and therefore radiative-convective boundaries, within the planet atmosphere, we cannot consistently calculate the time evolution of different atmospheres if we evaluate our cooling Equation (\ref{eq:coolingglobal}) at the RCB, as we do in our standard model. We choose to evaluate the cooling time at the Bondi radius instead (since our cooling model applies at any radius $R$, see section \S\ref{struct}). %We note that our choice of $R$ does not change the estimate of the atmosphere evolution time, to order of magnitude, since the additional luminosity generated in all radiative regions is negligible for our opacity choice (also see Paper I). 

%\subsection{Outer Boundary Effects}

%\textbf{Reemphasize the fact that the atmosphere structure is determined by your outer boundary conditions: T_{out}, P_{out}, R_{out} $\rightarrow$ explore the separate effects of pressure, temperature and a (since the Hill radius is determined by a); show how temperature is the strongest effect. }

\section{Critical Core Mass}
\label{critical}

%\textbf{Define what it is (refer again to paper I also); show the Mcrit vs a for fixed disk life plot, for both real EOS, gamma 7/5 and gamma 5/3; justify the differences in terms of the EOS effects from 4.2; show that using a real EOS makes a significant difference to the results; however, the core masses we get are still doable. Again, a lot of text below will be used but needs reorganizing/rephrasing.}

%\textbf{Define what it is (also refer to paper I) and emphasize how it's different from $M_{crit}$ in standard calculations. Define the crossover mass and crossover time.}

In this section we put together the results obtained in Sections \S\ref{EOSeffects}  and \S\ref{sec:opacity}, and determine the minimum core mass, $M_{\rm crit}$, to initiate runaway gas accretion during a typical protoplanetary disk lifetime, $t= 3$ Myr. As in Paper I, we quantify the runaway accretion time $t_{\rm run}$ as the time at which the atmosphere growth timescale $M_{\rm atm}/\dot{M}$ drops to 10\% of its maximum value (see Paper I for details). 

%We first explore the dependence of $t_{\rm run}$ on the core mass for a fixed semimajor axis. We then determine the critical core mass $M_{\rm crit}$ to form a giant planet from a gas composed of a realistic hydrogen-helium mixture, and we compare this with the results from Paper I for an ideal diatomic gas. Finally, we determine $M_{\rm crit}$ under more realistic opacity assumptions.

%\subsection{Crossover Time as a Function of Core Mass}
%\label{tvsM}

%\textbf{t vs. M at fixed distance, similar to the plot from paper I. Compare scalings.}

\begin{figure}[h!]
\centering
\includegraphics[width=0.5\textwidth]{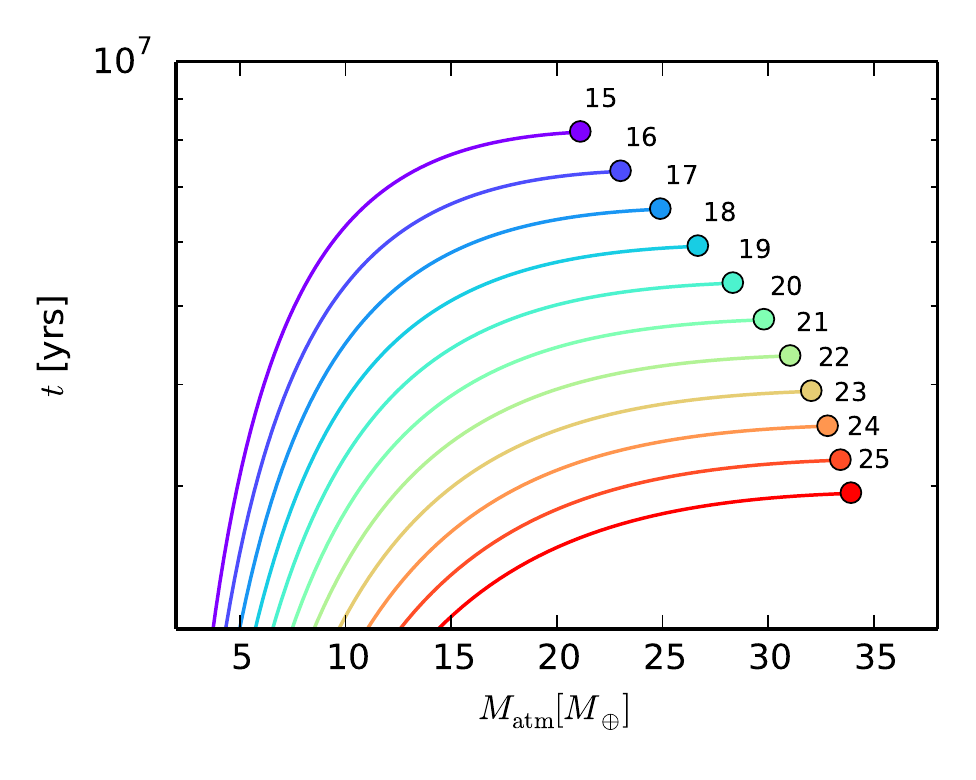}
%%\vspace{-0.5in}
\caption{Elapsed time as a function of atmosphere mass, for cores with fixed masses between $15 M_{\oplus}$ and $25 M_{\oplus}$ at $a=10$ AU in our fiducial disk, for a realistic EOS with an equilibrium ortho-to-para ratio and standard ISM opacity. The circles mark the runaway growth time. The numbers label the core mass in Earth masses. A larger core mass results in a lower $t_{\rm run}$.}% and a higher $M_{\rm atm}/M\co$ at runaway.}
\label{fig:tvsMplot}
\end{figure}

Figure \ref{fig:tvsMplot} displays the time evolution and the runaway growth time for atmospheres forming around cores with masses between 15 $M_{\oplus}$ and 25 $M_{\oplus}$ at $a=10$ AU in our fiducial disk, for a realistic EOS with an equilibrium ortho-to-para ratio and standard ISM opacity. Higher mass cores have shorter $t_{\rm run}$, consistent with the results of Paper I. We also note that $M_{\rm atm}$ at $t_{\rm run}$ is larger for the realistic EOS than for the polytropic EOS considered in Paper I, for the same core mass.

%\subsection{Critical Core Mass}
%\label{Mcrit}

%\textbf{Mcrit vs. a plot, realistic EOS and polytrope. Discuss the larger critical core mass for the real EOS in light of the effects from section 4.}

\begin{figure}[h!]
\centering
\includegraphics[width=0.5\textwidth]{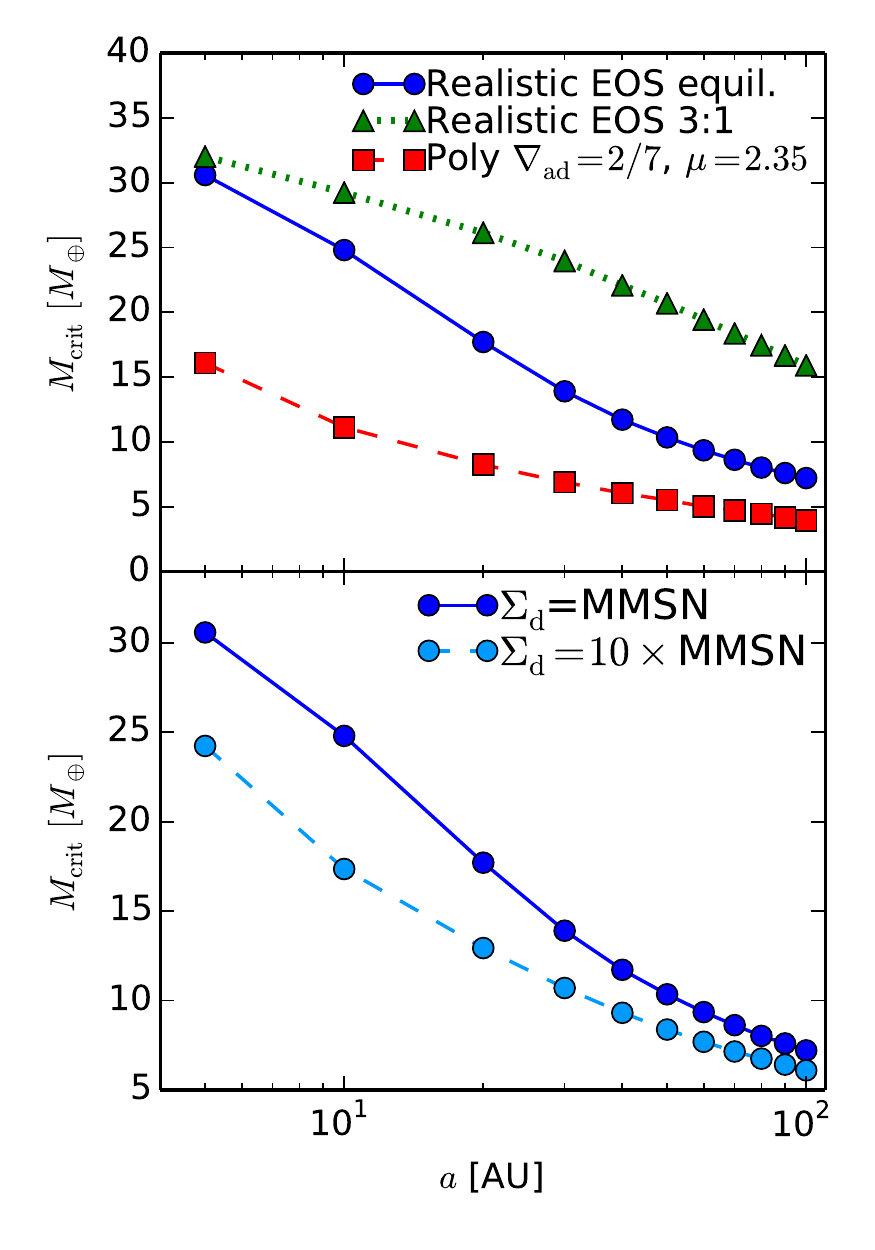}
%%\vspace{-0.5in}
\caption{Top panel: The minimum core mass for an atmosphere to initiate runaway gas accretion within the lifetime of a typical protoplanetary disk $t \sim 3$ Myrs as a function of semimajor axis, for a realistic hydrogen-helium mixture and a standard ISM opacity. The results of Paper I for an ideal diatomic gas are plotted for comparison. The realistic EOS yields core masses larger by a factor of $\sim$2 when compared to the polytrope, for an equilibrium ortho-to-para ratio. The critical core mass is $\sim$$2-4$ times larger than the polytrope case for a fixed 3:1 ratio between the $H_2$ spin isomers. The increase is more pronounced at larger stellocentric distances. Bottom panel: Critical core mass as a function of semimajor axis for a disk gas surface density 10 times larger than that of our fiducial disk. A larger $\Sigma_{\rm d}$ reduces $M_{\rm crit}$ by $\sim$$15-25 \%$.}
\label{fig:Mvsaplot}
\end{figure}

Figure \ref{fig:Mvsaplot}, upper panel, displays $M_{\rm crit}$
%(e.g., \citealt{jay99}), 
for a gas described by a realistic EOS and an ISM dust opacity. The results of Paper I for an ideal diatomic gas are plotted for comparison. When compared to an ideal gas polytrope, the inclusion of realistic EOS effects increases $M_{\rm crit}$ by a factor of $\sim$2 if the $H_2$ spin isomers are in equilibrium, and by a factor of $\sim$$2-4$ for a fixed 3:1 ortho-to-para ratio. This latter increase is more significant at larger stellocentric distances. In Figure \ref{fig:Mvsaplot}, bottom panel, we compare our results with those for a disk with a gas surface density an order of magnitude larger than $\Sigma_{\rm d}$ of our fiducial disk (see Equation \ref{eq:diska}), and find that $M_{\rm crit}$ reduces by $\sim$15-25$\%$.

 %As such, non-ideal effects substantially increase the core mass needed to form a giant planet  before the dissipation of the protoplanetary disk.   

%Figure \ref{fig:Mvsaplot} shows the critical core mass for a massive atmosphere to form during a typical lifetime of a protoplanetary disk $t=3$ Myrs 
%(e.g., \citealt{jay99}), 
%for a gas described by a realistic EOS and an ISM dust opacity. The results of Paper I for an ideal diatomic gas are plotted for comparison. The inclusion of realistic EOS effects increases $M_{\rm crit}$ by more than a factor of two when compared to an ideal gas polytrope. %As such, non-ideal effects substantially increase the core mass needed to form a giant planet  before the dissipation of the protoplanetary disk.   

\begin{figure}[h!]
\centering
\includegraphics[width=0.5\textwidth]{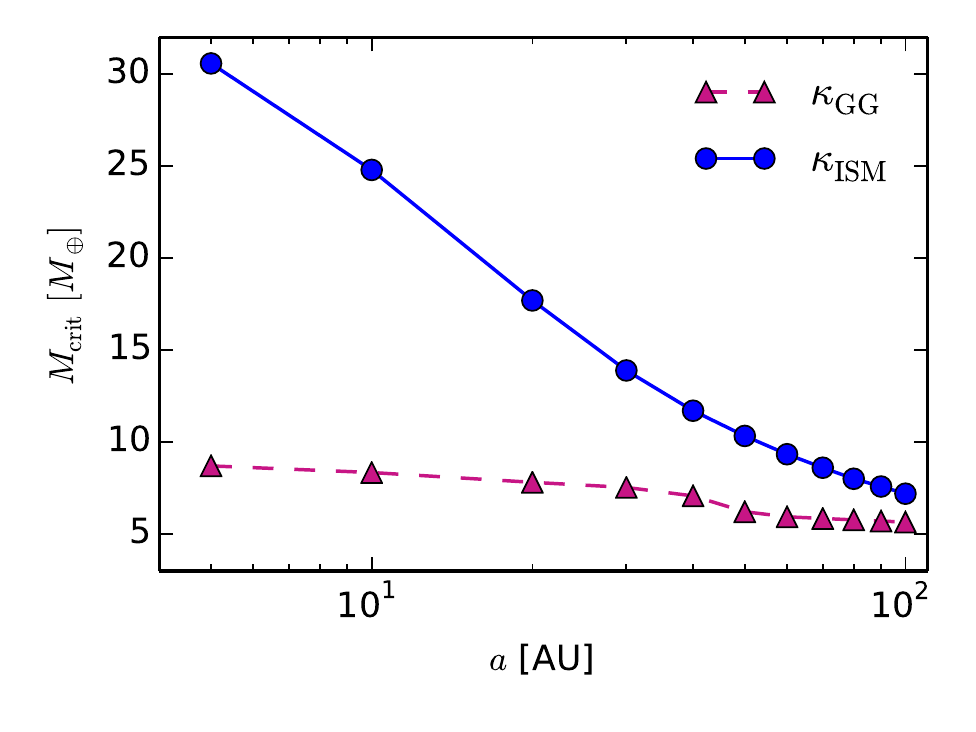}
%%\vspace{-0.5in}
\caption{Critical core mass as a function of semimajor axis for a realistic EOS with an equilibrium ortho-to-para ratio and radiative opacities that account for grain growth (purple triangles, with $p=3.5$ and $a_{\rm max}=1$ cm; see text for details). The critical core mass is lower than it would be if dust grains had an ISM-like size distribution (blue circles).}
\label{fig:Mcritvsagg}
\end{figure}

Figure \ref{fig:Mcritvsagg} shows $M_{\rm crit}$ as a function of semimajor axis, for a realistic EOS with an equilibrium ortho-to-para ratio and grain growth opacity with a size distribution given by Equation (\ref{eq:graindistr}) with $p=3.5$ and maximum particle size $s_{\rm max}=1$ cm. The critical core mass is lower than in the standard interstellar opacity case, and less sensitive to location in the disk. Location primarily affects the atmosphere through the opacity in the outer envelope, which depends on disk temperature. For the simplified analytic model developed in Paper I, we approximated $t_{\rm run}$ by  $t_{\rm co}$, the time when $M_{\rm atm}=M\co$, and found that $t_{\rm co} \sim T_{\rm d}^{\beta+1/2}$, with $\beta$ the power-law exponent in Equation (\ref{eq:opacitylaw}). Opacity is less sensitive to temperature variations for larger grains and has an almost flat profile (see Figure \ref{fig:opacity}), which results in $\beta \ll 1$ and a much weaker temperature (and therefore semimajor axis) dependence of $M_{\rm crit}$, as seen in Figure \ref{fig:Mcritvsagg}. Moreover, grain growth reduces the absolute value of the opacity, which also lowers $M_{\rm crit}$. % As noted in Section \S\ref{sec:opacity}, the critical core mass may be significantly lower if coagulation is taken into account, i.e. $p=2.5$.

\begin{figure}[h!]
\centering
\includegraphics[width=0.5\textwidth]{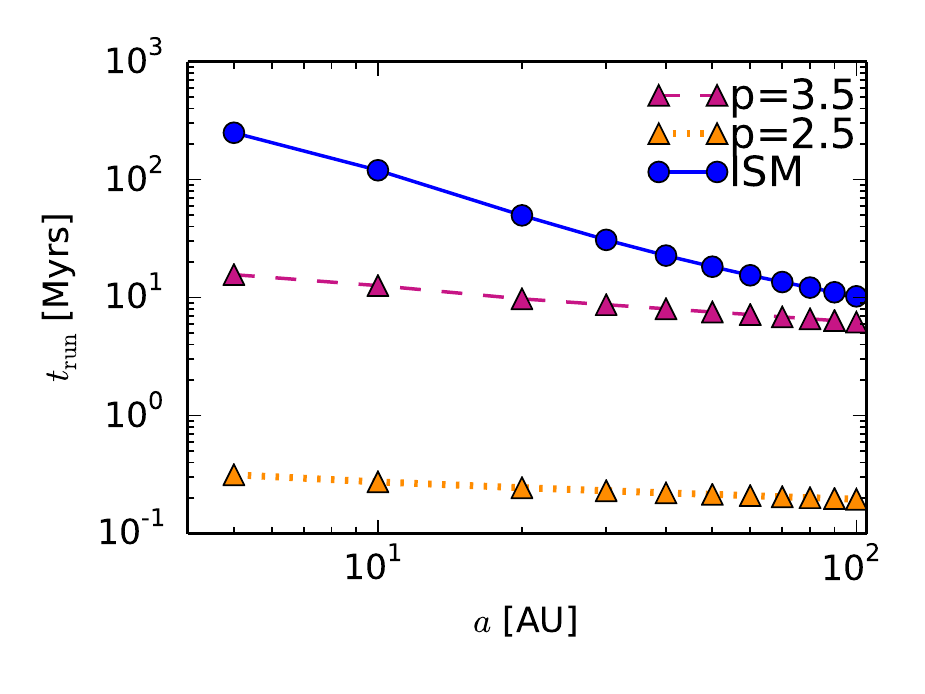}
%%\vspace{-0.5in}
\caption{Runaway accretion times for a realistic EOS with an equilibrium ortho-to-para ratio and different grain size distributions, for an atmosphere forming around a core with $M\co=4 M_{\oplus}$. The lines marked by purple and orange triangles have grain growth opacities with $a_{\rm max}=1$ cm, and $p=3.5$ and $p=2.5$, respectively (see text for details). The blue circle line has an ISM power-law opacity. The runaway accretion time is more than one order of magnitude lower when coagulation is accounted for, i.e. $p=2.5$.}
\label{fig:p25p35}
\end{figure}

In Equation (\ref{eq:graindistr}), the coefficient $p=3.5$ corresponds to a standard collisional cascade. If coagulation is taken into account, the exponent $p$ can be approximated as $p=2.5$ \citep{dalessio01}. This results in a flatter and significantly lower opacity (see \App{radwindow}), which may substantially reduce $M_{\rm crit}$. However, we have found that our model breaks down for low core masses ($M\co \lesssim 3 M_{\oplus}$) under our assumption of constant luminosity in the outer radiative layer. Figure \ref{fig:p25p35} shows the runaway accretion time $t_{\rm run}$ as a function of semimajor axis for the lowest core mass for which our model is valid, $M\co=4 M_{\oplus}$. The runaway accretion time is more than one order of magnitude lower for $p=2.5$, which implies that $M_{\rm crit}$ may be, in fact, significantly lower than presented in Figure \ref{fig:Mcritvsagg}. In other words, grain growth can yield critical core masses up to an order of magnitude lower than in the case where interstellar opacities are used.

In summary, we have found $M_{\rm crit}\sim 30 M_{\oplus}$ at 5 AU, steadily decreasing to $\sim$$7 M_{\oplus}$ at 100 AU, for a realistic EOS with the $H_2$ spin isomers in equilibrium and interstellar opacity. For a fixed 3:1 ortho-to-para ratio and interstellar opacity, $M_{\rm crit}$ is $\sim$$32 M_{\oplus}$ at 5 AU and decreases to $\sim$$15 M_{\oplus}$ at 100 AU. For a grain growth opacity with a size distribution given by Equation (\ref{eq:graindistr}) with $p=3.5$ and an equilibrium ortho-to-para ratio, $M_{\rm crit}$ significantly drops to $\sim$$8 M_{\oplus}$ at 5 AU and $\sim$$5 M_{\oplus}$ at 100 AU. Accounting for coagulation (i.e., $p=2.5$) $M_{\rm crit}$ is less than $4M_\oplus$ and may be up to an order of magnitude smaller.

\section{Effects of Planetesimal Accretion}
\label{acc}

% In this scenario, the energy the envelope receives from planetesimals balances its luminosity. The core's atmosphere is in steady state, and  $M_{\rm crit}$ is uniquely determined for a fixed planetesimal accretion rate, $\dot{M_{\rm c}}$, and a set of disk conditions. For high $\dot{M_{\rm c}}$, the resulting $M_{\rm crit}$ needed to grow an atmosphere during the disk lifetime is too large, which has led to the belief that core accretion does not work at large separations. 

This study considers protoplanets with fully formed cores for which planetesimal accretion is negligible and KH contraction dominates the luminosity evolution of the atmosphere. This approach contrasts with that of models which assume high planetesimal accretion rates and find that the atmosphere is in steady state and solely heated due to accretion of solids. In both cases, as the envelope and core become comparable in mass, hydrostatic balance no longer holds and runaway gas accretion commences. For fast accretion, $M_{\rm crit}$ is uniquely defined as the maximum core mass for which the atmosphere is still in hydrostatic equilibrium for a fixed planetesimal accretion rate and a set of disk conditions. In this section we compare our results for $M_{\rm crit}$ to analogous results from steady-state fast planetesimal accretion calculations. We discuss the core accretion rates that are necessary for our regime to be valid in \S\ref{raf1}. In \S\ref{raf3}, we estimate core growth at the maximum rate for which the KH regime is valid, and show it is negligible over the timescale on which the atmosphere evolves. Finally, we compare our results with those assuming fast planetesimal accretion in \S\ref{raf2}.

 %In this section we investigate the core accretion rates that are necessary for our regime to be valid. We also discuss the conditions under which runaway gas accretion can be initiated due to the Kelvin-Helmholtz contraction of the atmosphere before it becomes critical due to planetesimal accretion.

\subsection{Planetesimal Accretion Rates}
\label{raf1}

Kelvin-Helmholtz contraction dominates an atmosphere's luminosity if  $L_{\mathrm{acc}} < L_{\rm{KH}}$, where $L_{\rm{acc}}$ is the accretion luminosity,

\begin{equation}
\label{eq:Lacc}
L_{acc}=G \frac{M_{\rm{c}} \dot{M_{\rm{c}}}}{R_{\rm{c}}},
\end{equation}
and $L_{\rm KH}$ is given by Equation (\ref{eq:coolingglobal}) with $L\co=\Gamma=0$. This condition is satisfied as long as the planetesimal accretion rate 

\begin{equation}
\label{eq:McdotKH}
\dot{M\co}<\dot{M}_{\rm c, KH} \equiv \frac{L_{\rm KH} R\co}{G M\co}.
\end{equation} 
To illustrate the magnitude of $\dot{M}_{\rm c, KH}$, we choose as a fiducial case an atmosphere forming at 30 AU and with a core mass of $10 M_{\oplus}$. Since analytic studies of critical core masses at high planetesimal accretion rates assume an ideal gas EOS, for ease of comparison we choose an ideal gas polytropic EOS with constant adiabatic gradient $\delad=2/7$ and mean molecular weight $\mu=2.35$ (see also Paper I). For this choice of parameters, the runaway accretion time is $t _{\rm run}\sim$ 1.4 Myrs, which is within the typical lifetime of a protoplanetary disk. We also estimate two reference accretion rates. The first one is the core accretion rate $\dot{M}_{\rm c, acc}$ needed to grow the core to $M\co=10 M_{\oplus}$ on the same timescale as our model atmosphere, $\tau=1.4$ Myrs:

\begin{equation}
\label{eq:Mcdot}
\dot{M}_{\rm{c,acc}}(M_{\rm{c}}) \sim \frac{M_{\rm{c}}}{\tau}.
\end{equation}
The second reference planetesimal accretion rate is $\dot{M}_{\rm c, Hill}$, a typically assumed planetesimal accretion rate for which the random velocities of the planetesimals are of the order of the Hill velocity around the protoplanetary core (for a review, see \citealt{goldreich04}). Following R06 (equation A1),

%This is the accretion rate at the boundary between the dispersion dominated and shear dominated regimes. 

\begin{equation}
\label{eq:MdotHill}
\dot{M}_{\rm{c,Hill}}=\Omega \Sigma_{\rm p} R\co R_{\rm H},
\end{equation}
where $\Sigma_{\rm p}$ is the surface density of solids, assumed to satisfy $\Sigma\di \approx 100 \Sigma_{\rm p}$ for a dust-to-gas ratio of 0.01.

\begin{figure}[h]
\centering
\includegraphics[width=0.5\textwidth]{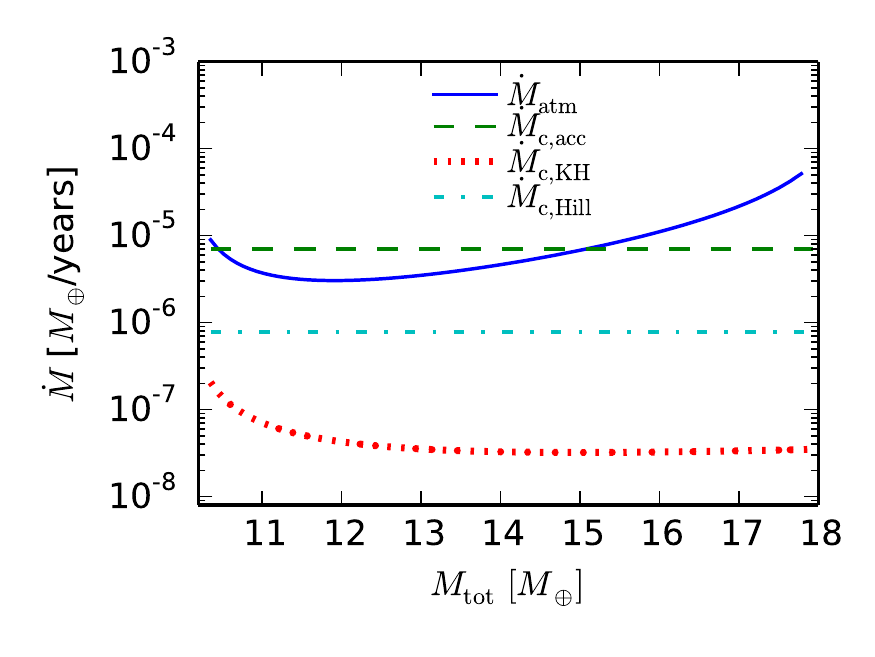}
%\vspace{-0.5in}
\caption{Various accretion rates for a planet forming at 30 AU and with a core mass $M\co=10 M_{\oplus}$, using a polyropic EOS and ISM power-law opacity. For this choice of parameters, the runaway accretion time is $t_{\rm run} \sim 1.4$ Myrs. The $\dot{M}_{\rm{atm}}$ (solid blue) curve represents the growth rate of the atmosphere as estimated by our model. The core accretion rate $\dot{M}_{\rm c, acc}$ (dashed green) necessary to grow the core on the timescale $t_{\rm run}$ is larger than $\dot{M}_{\rm c, KH}$ (dotted red), the maximum planetesimal accretion rate during KH contraction for which our regime is valid (see text). The frequently used planetesimal accretion rate $\dot{M}_{\rm c, Hill}$ (dashed-dotted light blue) for which the random velocity of the planetesimals is given by the Hill velocity due to the core (see text), also exceeds $\dot{M}_{\rm c, KH}$.This motivates our requirement  that planetesimal accretion must have slowed down after core growth for our model to be valid.}

%and $\dot{M}_{\rm{c,KH}}$ (dotted red) is the maximum planetesimal accretion rate during the gas contraction phase in order for our regime to be valid, i.e. $L_{\mathrm{acc}} < L_{\rm{KH}}$ (see text). For comparison, we plot the core accretion rate $\dot{M}_{\rm{c,acc}}$ (dashed green) necessary to grow the core on a timescale $\tau=t_{\rm run}$, and a frequently used planetesimal accretion rate $\dot{M}_{\rm{c,Hill}}$ (dash-dotted light blue) for which the random velocity of the planetesimals is given by the Hill velocity due to the core (see text). We note that $\dot{M}_{\rm c, KH}$ and $\dot{M}_{\rm c, Hill}$ are both lower than $\dot{M}_{\rm c, acc}$, motivating our requirement  that planetesimal accretion must have slowed down after core growth for our model to be valid.}
\label{fig:accrates}
\end{figure}

Figure \ref{fig:accrates} shows that $\dot{M}_{\rm c,KH}$ is $\sim2-3$ orders of magnitude lower than $\dot{M}_{\rm c, acc}$. Had the core accreted planetesimals at the $\dot{M}_{\rm c, KH}$ rate since it started forming, it could not have grown large enough to attract an atmosphere within a typical disk lifetime. Our model requires that the planetesimal accretion rate is initially large during core growth, then significantly reduces as the gaseous envelope accumulates, as suggested by, e.g., \citet{pollack96}. This is a plausible scenario: the core's feeding zone may be depleted of solids if it is not refilled by radial drift of planetesimals through the nebula, or the core may form in the inner part of the disk and later be scattered outwards by other giants in the system \citep{ida13}. 

%or the planet's feeding zone could have been depleted of solids due to a giant neighbor.  

\subsection{Core Growth during KH Contraction}
\label{raf3}

Planetesimal accretion during the KH contraction phase of atmosphere growth at a rate $\dot{M}_{\rm c}<\dot{M}_{\rm{c,KH}}$  cannot alter the core mass enough to affect the time evolution of the atmosphere. We can quantitatively estimate the maximum increase in core mass as 

\begin{equation}
\label{eq:cminc}
\Delta M_{\rm{c}} = \int_0^{t_{\rm{run}}} \dot{M_{\rm{c}}} dt \approx \sum_i \dot{M_{\rm{c}}}_i \Delta t_i,
\end{equation}
 
 \noindent where the accretion rate $ \dot{M_{\rm{c}}}_i $ is given by 
 
 \begin{equation}
 \label{eq:Mdotexp}
 \dot{M_{\rm{c}}}_i =\frac{L_i R\co}{G M_{\rm{c}}} 
 \end{equation}
 
 \noindent from Equation (\ref{eq:Lacc}), with $L_i$ the luminosity of the atmosphere at time $t_i$ in our model. For $M_{\rm{c}}=10 M_{\oplus}$, we find $\Delta M_{\rm{c}} \approx 0.05 M_{\oplus} \ll 10 M_{\oplus}$. Core growth is negligible in our regime, and the time evolution of the atmosphere is thus insensitive to core mass changes at a rate imposed by the assumption that $L_{\rm{acc}}<L_{\rm{KH}}$.

 %It is easy to see that  $\dot{M}_{\rm{c,typical}}$ is more than one order of magnitude lower than the gas accretion rate of our model atmosphere $\dot{M}_{\rm{atm}}$, and lower than the core accretion rate $\dot{M}_{\rm{c,acc}}$ needed to grow the core and the atmosphere at the same time within the disk life time. As such, the formation of a giant planet by growing the core first, then letting the atmosphere cool is faster than growing the core and the atmosphere at the same time at a steady planetesimal accretion rate.

\subsection{Comparison with Steady-State Results}
\label{raf2}

We compare our results for $M_{\rm crit}$ with those of studies that assume large planetesimal accretion rates.  In principle, the disk lifetime could be short enough that our calculated $M_{\rm crit}$ 
%calculated using KH contraction in the absence of planetesimal accretion 
could exceed the $M_{\rm crit}$ estimated for a steady-state, accretion-heated core.  This prospect is not self-contradictory since at higher luminosity, an atmosphere evolves more quickly and hence reaches steady state on a timescale shorter than our calculated KH contraction time.   
%If the $M_{\rm crit}$ found by these studies were lower than the $M_{\rm crit}$ in our regime of negligible solids accretion, the atmosphere would undergo runaway gas accretion before KH contraction became dominant, and our regime would not be relevant. 
We show here that disk lifetimes are long enough that this is not the case.  Our model yields lower core masses than those found when fast planetesimal accretion is considered. 

The critical core mass is larger for higher planetesimal accretion, as additional heating increases the core mass required for collapse. As such, if atmosphere collapse does not occur for the lowest value of $\dot{M}_{\rm c, KH}$ over the course of the atmosphere's growth, then it can only occur in the KH dominated regime. %In what follows we calculate the critical core mass $M_{\rm crit, KH}$ corresponding to $\min(\dot{M}_{\rm c, KH})$ and show that it is higher that the critical core mass we determined in the KH dominated regime. 

In order to estimate the critical core mass $M_{\rm crit, KH}$ corresponding to planetesimal accretion at the rate $\dot{M}_{\rm c, KH}$, we use the results of R06 for low luminosity atmospheres forming in the outer disk ($\gtrsim2-5$ AU). %, consistent with our region of interest. 
R06 assumes an ideal gas polytropic EOS and an opacity lower than that of the ISM (see Equation \ref{eq:opacitylaw}). For comparison, we calculate $M_{\rm crit}$ for an ideal gas polytrope and an opacity, $\kappa$, given by Equation \ref{eq:opacitylaw} with $F_{\kappa}$ reduced by a factor of 100.  This choice is comparable to the opacity law used by R06.\footnote{The power-law opacity of R06 is scaled to the (semimajor axis dependent) disk temperature, while our opacity is scaled to an absolute reference temperature. We thus cannot directly use the R06 opacities for our comparison.} %(see Paper I). 

\begin{figure}[h]
\centering
\includegraphics[width=0.5\textwidth]{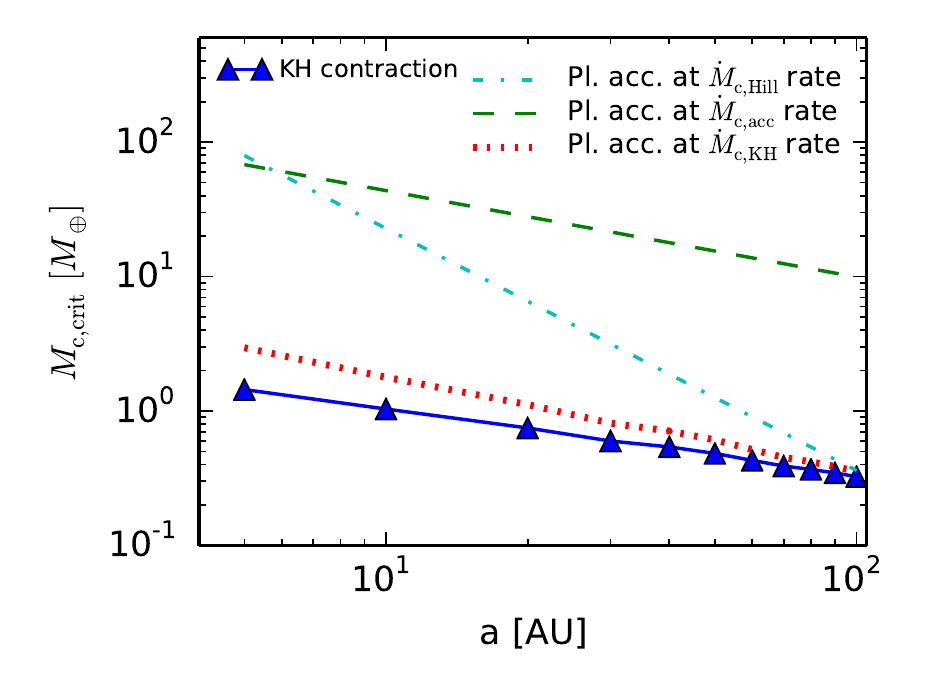}
%\vspace{-0.5in}
\caption{Comparison between the critical core mass $M_{\rm{crit, KH}}$ given significant planetesimal accretion and the critical core mass when gas contraction dominates, for a polytropic EOS and an ISM opacity reduced by a factor of 100. Our results yield lower core masses than in the fast planetesimal accretion case (e.g., \citealt{rafikov06}). The critical core mass corresponding to $\dot{M}_{\rm c, Hill}$ and $\dot{M}_{\rm c, acc}$ from Figure \ref{fig:accrates} is plotted for comparison.}
\label{fig:raf2}
\end{figure}

Following R06, we find that the critical core mass when accretion luminosity dominates the evolution of the atmosphere can be expressed as

\begin{equation}
\label{eq:critraf}
M_{\rm{crit, KH}} \sim \Big[\frac{\min[\dot{M}_{\rm c, KH}(M_{\rm{c}})]}{64 \pi^2 C} \frac{\kappa}{\sigma G^3} \frac{1}{R\co M\co^{1/3}} \Big(\frac{k_B}{\mu m_p}\Big)^4\Big]^{3/5},
\end{equation}
where $C$ is an order unity constant depending on the adiabatic gradient and disk properties (see R06, Equation B3). From Equation (\ref{eq:McdotKH}), the accretion rate $\dot{M}_{\rm c, KH}$ depends on the core mass $M\co$. We find $M_{\rm crit, KH}$ numerically by setting $M\co=M_{\rm crit, KH}$ on the right-hand side of Equation (\ref{eq:critraf}). The result is displayed in Figure \ref{fig:raf2}; the critical core mass corresponding to planetesimal accretion at the rates displayed in Figure \ref{fig:accrates} is displayed for comparison.  

The critical core mass in the regime where KH contraction dominates is smaller than in the case in which planetesimal accretion dominates the evolution of the atmosphere. This leads to two conclusions:

\begin{enumerate}
\item Planetesimal accretion can be safely ignored in our regime.
\item Giant planets can form from smaller cores if planetesimal accretion significantly reduces during atmosphere growth. 
\end{enumerate}

%As additional heating due to planetesimals limits the atmosphere's ability to cool, our result represents a true minimum for the core mass needed to form a gas giant during the lifetime of the protoplanetary disk.

% First, we confirm that planetesimal accretion can be safely ignored in our regime of interest. Secondly, this comparison tells us that the core mass needed to form a giant planet is smaller if the core forms first, then accretes a massive envelope, than in the case where the core and atmosphere grown simultaneously in a high planetesimal accretion regime. Moreover, our result represents a true, absolute minimum on the core mass needed to form a giant planet during the lifetime of the protoplanetary disk, as our core no longer grows.

 \section{Summary}
 \label{conclusions}
 
 %OK, very interesting.  Clearly the text (including abstract and summary I think) needs to be modified to include this result.  The headline for this result seems to be: at very low temperatures, the choice of ortho/para ratio makes as big a difference as just using a simple polytrope vs. a real EOS!!  This is consistent with our knowledge that spin effects are more important than dissociation once the spin levels are no longer thermally populated ($T_disk \lesssim T_spin \sim 90$K).  And indeed the behavior for different ortho/para cases is similar at higher temperature, though we don?t go to high enough temp $T > few x T_spin$ for ortho/para ratio to become completely irrelevant.

 In this paper we study the formation of giant planets embedded in a gas disk. We consider atmospheric evolution around fully grown cores and determine the minimum (critical) core mass, $M_{\rm crit}$, required to form a gas giant during the typical lifetime of a protoplanetary disk. We improve the model developed in  \citet[hereafter Paper I]{piso14} by including realistic equation of state (EOS) tables and dust opacities. 
 
 For a realistic EOS with the molecular hydrogen ($H_2$) spin isomers in thermal equilibrium, and grain growth opacity with maximum particle size $s_{\rm max}=1$ cm and a power-law size distribution (\ref{eq:graindistr}) with $p=3.5$, $M_{\rm crit}$ is $\sim$$8 M_{\oplus}$ at 5 AU in our fiducial disk and drops to $\sim$$5 M_{\oplus}$ at 100 AU. The realistic EOS and grain growth opacity have two competing effects on $M_{\rm crit}$: 
 
 \begin{enumerate}
 \item Realistic EOS effects increase $M_{\rm crit}$ by a factor of $\sim$$2$ when compared to the ideal gas polytrope, for an equilibrium ratio of ortho- and parahydrogen. If the $H_2$ spin isomers are in a fixed 3:1 ratio, $M_{\rm crit}$ increases by a factor of $\sim$$2-4$ when compared to the polytrope. This increase is most significant at larger stellocentric distances, where disk temperatures are lower than the peak ortho-to-para conversion temperature of $\sim$$50$ K.
  
 \item Grain growth opacities decrease $M_{\rm crit}$ by a factor of $\sim$$3.5$ at 5 AU and by a factor of $\sim$$1.2$ at 100 AU, when compared to ISM opacities, for a particle distribution given by the power-law (\ref{eq:graindistr}) with $p=3.5$ and a maximum particle size of 1 cm. The critical core mass is less sensitive to the location in the disk when realistic opacities are used. If $p=2.5$, an approximation for coagulation,  $M_{\rm crit}$ further reduces by up to an order of magnitude. 
 \end{enumerate}

 %interstellar grain opacity, the critical core mass is $\sim$$20 M_{\oplus}$ at 5 AU and drops to $\sim$$6 M_{\oplus}$ in our fiducial disk. These results are more than twice as large than those calculated in Paper I for a polytropic EOS, and bring the critical core mass to $\sim$$10 M_{\oplus}$, the value typically quoted in many core accretion studies. When realistic opacities that include grain growth are used, the critical core mass is significantly lower, i.e. $\sim$$6 M_{\oplus}$ at 5 AU and $\sim$$4 M_{\oplus}$ at 100 AU. Different assumptions about the grain size distribution (e.g., that take into account coagulation) may further reduce the critical core mass.
 %\vspace{2in}
 
 In our core accretion models, the dissociation of molecular hydrogen slows the atmospheric cooling and gas accretion rate.  This finding is somewhat surprising because $H_2$ dissociation can trigger the gravitational collapse of protostellar or planetary mass gas clouds (\citealt{bodenheimer80}, \citealt{inutsuka12}).  The presence of a massive solid core is the key factor that prevents dissociation from inducing a global collapse in our models.
 
%%We note that before the onset of runaway gas accretion, hydrogen dissociation deep in the envelope slows  the contraction of a core's atmosphere. When runaway accretion has increased the planet's mass substantially, dissociation can have the opposite effect, triggering collapse \citep{bodenheimer80}. 

%Dissociation is known to trigger the collapse of growing stars  \citep{larson69}. % on higher mass objects, such as  growing stars and collapsing planets forming stars can trigger the collapse of higher mass objects (\citealt{larson69}, \citealt{bodenheimer80}), .  deep in the envelope slows atmospheric contraction. This result contrasts with the effect of dissociation on star formation, where it can trigger dynamical instabilities and the collapse of higher mass objects (\citealt{larson69}, \citealt{bodenheimer80}). 
 
Our results yield lower core masses than analogous results that consider high planetesimal accretion rates for which the core and atmosphere grow simultaneously. It is thus possible to form a giant planet from a smaller core if the core grows first, then the accretion rate of solids is reduced and a gaseous envelope is accumulated. Moreover, since additional heat sources such as planetesimal accretion limit the ability of the atmosphere to cool and undergo Kelvin-Helmholtz contraction, our results represent a true minimum on the core mass needed to form a giant planet during the typical lifetime of a protoplanetary disk.

\acknowledgements{We thank the referee for a helpful report. We thank Sean Andrews for providing us with the \citet{dalessio01} opacity tables. We also thank Phil Arras, Eugene Chiang and Steven Cranmer for useful conversations.  ANY acknowledges support from {\it NASA} {\it Astrophysics Theory Program} and  {\it Origins of Solar Systems Program}  through grant NNX10AF35G. AP and RMC acknowledge support from the Brinson Foundation.}

\bibliographystyle{apj}
\bibliography{refs}

\begin{thebibliography}{69}
\expandafter\ifx\csname natexlab\endcsname\relax\def\natexlab#1{#1}\fi

\bibitem[{{Alibert} {et~al.}(2005){Alibert}, {Mordasini}, {Benz}, \&
  {Winisdoerffer}}]{alibert05}
{Alibert}, Y., {Mordasini}, C., {Benz}, W., \& {Winisdoerffer}, C. 2005, \aap,
  434, 343

\bibitem[{{Arras} \& {Bildsten}(2006)}]{arras06}
{Arras}, P. \& {Bildsten}, L. 2006, \apj, 650, 394

\bibitem[{{Bai} \& {Goodman}(2009)}]{bai09}
{Bai}, X.-N. \& {Goodman}, J. 2009, \apj, 701, 737

\bibitem[{{Beckwith} \& {Sargent}(1991)}]{beckwith91}
{Beckwith}, S.~V.~W. \& {Sargent}, A.~I. 1991, \apj, 381, 250

\bibitem[{{Beckwith} {et~al.}(1990){Beckwith}, {Sargent}, {Chini}, \&
  {Guesten}}]{beckwith90}
{Beckwith}, S.~V.~W., {Sargent}, A.~I., {Chini}, R.~S., \& {Guesten}, R. 1990,
  \aj, 99, 924

\bibitem[{{Bell} \& {Lin}(1994)}]{bell94}
{Bell}, K.~R. \& {Lin}, D.~N.~C. 1994, \apj, 427, 987

\bibitem[{Blanksby \& {Ellison}(2003)}]{blanksby03}
Blanksby, S.~J. \& {Ellison}, G.~B. 2003, Acc. Chem. Res., 36, 255

\bibitem[{{Bodenheimer} {et~al.}(1980){Bodenheimer}, {Grossman}, {Decampli},
  {Marcy}, \& {Pollack}}]{bodenheimer80}
{Bodenheimer}, P., {Grossman}, A.~S., {Decampli}, W.~M., {Marcy}, G., \&
  {Pollack}, J.~B. 1980, Icarus, 41, 293

\bibitem[{{Bodenheimer} \& {Pollack}(1986)}]{boden86}
{Bodenheimer}, P. \& {Pollack}, J.~B. 1986, Icarus, 67, 391

\bibitem[{{Boley} {et~al.}(2007){Boley}, {Hartquist}, {Durisen}, \&
  {Michael}}]{boley07}
{Boley}, A.~C., {Hartquist}, T.~W., {Durisen}, R.~H., \& {Michael}, S. 2007,
  \apjl, 656, L89

\bibitem[{{Chiang} \& {Youdin}(2010)}]{chiang10}
{Chiang}, E. \& {Youdin}, A.~N. 2010, Annual Review of Earth and Planetary
  Sciences, 38, 493

\bibitem[{{Conrath} \& {Gierasch}(1984)}]{conrath84}
{Conrath}, B.~J. \& {Gierasch}, P.~J. 1984, Icarus, 57, 184

\bibitem[{{D'Alessio} {et~al.}(2001){D'Alessio}, {Calvet}, \&
  {Hartmann}}]{dalessio01}
{D'Alessio}, P., {Calvet}, N., \& {Hartmann}, L. 2001, \apj, 553, 321

\bibitem[{{D'Angelo} \& {Bodenheimer}(2013)}]{dangelo13}
{D'Angelo}, G. \& {Bodenheimer}, P. 2013, \apj, 778, 77

\bibitem[{{D'Angelo} {et~al.}(2011){D'Angelo}, {Durisen}, \&
  {Lissauer}}]{dangelo11}
{D'Angelo}, G., {Durisen}, R.~H., \& {Lissauer}, J.~J. {Giant Planet
  Formation}, ed. S.~{Piper}, 319--346

\bibitem[{{Dohnanyi}(1969)}]{dohnanyi69}
{Dohnanyi}, J.~S. 1969, \jgr, 74, 2531

\bibitem[{{Farkas}(1935)}]{farkas35}
{Farkas}, A. 1935, {Orthohydrogen, Parahydrogen and Heavy Hydrogen}

\bibitem[{{Fouchet} {et~al.}(2003){Fouchet}, {Lellouch}, \&
  {Feuchtgruber}}]{fouchet03}
{Fouchet}, T., {Lellouch}, E., \& {Feuchtgruber}, H. 2003, Icarus, 161, 127

\bibitem[{{Fukutani} \& {Sugimoto}(2013)}]{fukutani13}
{Fukutani}, K. \& {Sugimoto}, T. 2013, Progress In Surface Science, 88, 279

\bibitem[{{Glassgold} {et~al.}(1997){Glassgold}, {Najita}, \&
  {Igea}}]{glassgold97}
{Glassgold}, A.~E., {Najita}, J., \& {Igea}, J. 1997, \apj, 480, 344

\bibitem[{{Goldreich} {et~al.}(2004){Goldreich}, {Lithwick}, \&
  {Sari}}]{goldreich04}
{Goldreich}, P., {Lithwick}, Y., \& {Sari}, R. 2004, \araa, 42, 549

\bibitem[{{Graboske} {et~al.}(1969){Graboske}, {Harwood}, \&
  {Rogers}}]{graboske69}
{Graboske}, H.~C., {Harwood}, D.~J., \& {Rogers}, F.~J. 1969, Physical Review,
  186, 210

\bibitem[{{Hori} \& {Ikoma}(2010)}]{hori10}
{Hori}, Y. \& {Ikoma}, M. 2010, \apj, 714, 1343

\bibitem[{{Hubickyj} {et~al.}(2005){Hubickyj}, {Bodenheimer}, \&
  {Lissauer}}]{hubickyj05}
{Hubickyj}, O., {Bodenheimer}, P., \& {Lissauer}, J.~J. 2005, Icarus, 179, 415

\bibitem[{{Huestis}(2008)}]{huestis08}
{Huestis}, D.~L. 2008, \planss, 56, 1733

\bibitem[{{Ida} {et~al.}(2013){Ida}, {Lin}, \& {Nagasawa}}]{ida13}
{Ida}, S., {Lin}, D.~N.~C., \& {Nagasawa}, M. 2013, \apj, 775, 42

\bibitem[{{Ikoma} {et~al.}(2000){Ikoma}, {Nakazawa}, \& {Emori}}]{ikoma00}
{Ikoma}, M., {Nakazawa}, K., \& {Emori}, H. 2000, \apj, 537, 1013

\bibitem[{{Inutsuka}(2012)}]{inutsuka12}
{Inutsuka}, S.-i. 2012, Progress of Theoretical and Experimental Physics, 2012,
  010000

\bibitem[{{Jayawardhana} {et~al.}(1999){Jayawardhana}, {Hartmann}, {Fazio},
  {Fisher}, {Telesco}, \& {Pi{\~n}a}}]{jay99}
{Jayawardhana}, R., {Hartmann}, L., {Fazio}, G., {Fisher}, R.~S., {Telesco},
  C.~M., \& {Pi{\~n}a}, R.~K. 1999, \apjl, 521, L129

\bibitem[{{Kippenhahn} \& {Weigert}(1990)}]{kippenhahn90}
{Kippenhahn}, R. \& {Weigert}, A. 1990, {Stellar Structure and Evolution}

\bibitem[{{Kittel} {et~al.}(1981){Kittel}, {Kroemer}, \& {Landsberg}}]{kittel}
{Kittel}, C., {Kroemer}, H., \& {Landsberg}, P.~T. 1981, \nat, 289, 729

\bibitem[{{Langmuir}(1912)}]{langmuir12}
{Langmuir}, I. 1912, J. Am. Chem. Soc., 34, 860

\bibitem[{{Larson}(1969)}]{larson69}
{Larson}, R.~B. 1969, \mnras, 145, 271

\bibitem[{{Lique} {et~al.}(2012){Lique}, {Honvault}, \& {Faure}}]{lique12}
{Lique}, F., {Honvault}, P., \& {Faure}, A. 2012, \jcp, 137, 154303

\bibitem[{{Lique} {et~al.}(2014){Lique}, {Honvault}, \& {Faure}}]{lique14}
---. 2014, ArXiv e-prints, arXiv:1402.5292

\bibitem[{{Mandl}(1989)}]{mandl89}
{Mandl}, F. 1989, {Statistical Physics, 2nd Edition}

\bibitem[{{Marois} {et~al.}(2008){Marois}, {Macintosh}, {Barman}, {Zuckerman},
  {Song}, {Patience}, {Lafreni{\`e}re}, \& {Doyon}}]{marois08}
{Marois}, C., {Macintosh}, B., {Barman}, T., {Zuckerman}, B., {Song}, I.,
  {Patience}, J., {Lafreni{\`e}re}, D., \& {Doyon}, R. 2008, Science, 322, 1348

\bibitem[{{Milenko} {et~al.}(1997){Milenko}, {Sibileva}, \&
  {Strzhemechny}}]{milenko97}
{Milenko}, Y.~Y., {Sibileva}, R.~M., \& {Strzhemechny}, M.~A. 1997, Journal of
  Low Temperature Physics, 107, 77

\bibitem[{{Militzer} \& {Hubbard}(2013)}]{militzer13}
{Militzer}, B. \& {Hubbard}, W.~B. 2013, \apj, 774, 148

\bibitem[{{Mizuno} {et~al.}(1978){Mizuno}, {Nakazawa}, \& {Hayashi}}]{mizuno78}
{Mizuno}, H., {Nakazawa}, K., \& {Hayashi}, C. 1978, Progress of Theoretical
  Physics, 60, 699

\bibitem[{{Mordasini}(2014)}]{mordasini14b}
{Mordasini}, C. 2014, \aap, 572, A118

\bibitem[{{Mordasini} {et~al.}(2012){Mordasini}, {Alibert}, {Klahr}, \&
  {Henning}}]{mordasini12}
{Mordasini}, C., {Alibert}, Y., {Klahr}, H., \& {Henning}, T. 2012, \aap, 547,
  A111

\bibitem[{{Mordasini} {et~al.}(2014){Mordasini}, {Klahr}, {Alibert}, {Miller},
  \& {Henning}}]{mordasini14}
{Mordasini}, C., {Klahr}, H., {Alibert}, Y., {Miller}, N., \& {Henning}, T.
  2014, ArXiv e-prints

\bibitem[{{Movshovitz} {et~al.}(2010){Movshovitz}, {Bodenheimer}, {Podolak}, \&
  {Lissauer}}]{movshovitz10}
{Movshovitz}, N., {Bodenheimer}, P., {Podolak}, M., \& {Lissauer}, J.~J. 2010,
  Icarus, 209, 616

\bibitem[{{Nettelmann} {et~al.}(2012){Nettelmann}, {Becker}, {Holst}, \&
  {Redmer}}]{nettelmann12}
{Nettelmann}, N., {Becker}, A., {Holst}, B., \& {Redmer}, R. 2012, \apj, 750,
  52

\bibitem[{{Nettelmann} {et~al.}(2008){Nettelmann}, {Holst}, {Kietzmann},
  {French}, {Redmer}, \& {Blaschke}}]{nettelmann08}
{Nettelmann}, N., {Holst}, B., {Kietzmann}, A., {French}, M., {Redmer}, R., \&
  {Blaschke}, D. 2008, \apj, 683, 1217

\bibitem[{{Ormel}(2013)}]{ormel13}
{Ormel}, C.~W. 2013, \mnras, 428, 3526

\bibitem[{{Ormel}(2014)}]{ormel14}
---. 2014, \apjl, 789, L18

\bibitem[{{Pachucki} \& {Komasa}(2008)}]{pachucki08}
{Pachucki}, K. \& {Komasa}, J. 2008, \pra, 77, 030501

\bibitem[{{Papaloizou} \& {Nelson}(2005)}]{pn05}
{Papaloizou}, J.~C.~B. \& {Nelson}, R.~P. 2005, \aap, 433, 247

\bibitem[{{Papaloizou} \& {Terquem}(1999)}]{pap99}
{Papaloizou}, J.~C.~B. \& {Terquem}, C. 1999, \apj, 521, 823

\bibitem[{{Paxton} {et~al.}(2011){Paxton}, {Bildsten}, {Dotter}, {Herwig},
  {Lesaffre}, \& {Timmes}}]{paxton11}
{Paxton}, B., {Bildsten}, L., {Dotter}, A., {Herwig}, F., {Lesaffre}, P., \&
  {Timmes}, F. 2011, \apjs, 192, 3

\bibitem[{{Paxton} {et~al.}(2013){Paxton}, {Cantiello}, {Arras}, {Bildsten},
  {Brown}, {Dotter}, {Mankovich}, {Montgomery}, {Stello}, {Timmes}, \&
  {Townsend}}]{paxton13}
{Paxton}, B., {Cantiello}, M., {Arras}, P., {Bildsten}, L., {Brown}, E.~F.,
  {Dotter}, A., {Mankovich}, C., {Montgomery}, M.~H., {Stello}, D., {Timmes},
  F.~X., \& {Townsend}, R. 2013, \apjs, 208, 4

\bibitem[{{P{\'e}rez} {et~al.}(2012){P{\'e}rez}, {Carpenter}, {Chandler},
  {Isella}, {Andrews}, {Ricci}, {Calvet}, {Corder}, {Deller}, {Dullemond},
  {Greaves}, {Harris}, {Henning}, {Kwon}, {Lazio}, {Linz}, {Mundy}, {Sargent},
  {Storm}, {Testi}, \& {Wilner}}]{perez12}
{P{\'e}rez}, L.~M., {Carpenter}, J.~M., {Chandler}, C.~J., {Isella}, A.,
  {Andrews}, S.~M., {Ricci}, L., {Calvet}, N., {Corder}, S.~A., {Deller},
  A.~T., {Dullemond}, C.~P., {Greaves}, J.~S., {Harris}, R.~J., {Henning}, T.,
  {Kwon}, W., {Lazio}, J., {Linz}, H., {Mundy}, L.~G., {Sargent}, A.~I.,
  {Storm}, S., {Testi}, L., \& {Wilner}, D.~J. 2012, \apjl, 760, L17

\bibitem[{{Perez-Becker} \& {Chiang}(2011)}]{perez11}
{Perez-Becker}, D. \& {Chiang}, E. 2011, \apj, 727, 2

\bibitem[{{Piso} \& {Youdin}(2014)}]{piso14}
{Piso}, A.-M.~A. \& {Youdin}, A.~N. 2014, \apj, 786, 21

\bibitem[{{Pollack} {et~al.}(1996){Pollack}, {Hubickyj}, {Bodenheimer},
  {Lissauer}, {Podolak}, \& {Greenzweig}}]{pollack96}
{Pollack}, J.~B., {Hubickyj}, O., {Bodenheimer}, P., {Lissauer}, J.~J.,
  {Podolak}, M., \& {Greenzweig}, Y. 1996, Icarus, 124, 62

\bibitem[{{Pollack} {et~al.}(1985){Pollack}, {McKay}, \&
  {Christofferson}}]{pollack85}
{Pollack}, J.~B., {McKay}, C.~P., \& {Christofferson}, B.~M. 1985, Icarus, 64,
  471

\bibitem[{{Rafikov}(2006)}]{rafikov06}
{Rafikov}, R.~R. 2006, \apj, 648, 666

\bibitem[{{Rafikov}(2011)}]{rafikov11}
---. 2011, \apj, 727, 86

\bibitem[{{Saumon} {et~al.}(1995){Saumon}, {Chabrier}, \& {van
  Horn}}]{saumon95}
{Saumon}, D., {Chabrier}, G., \& {van Horn}, H.~M. 1995, \apjs, 99, 713

\bibitem[{{Semenov} {et~al.}(2003){Semenov}, {Henning}, {Helling}, {Ilgner}, \&
  {Sedlmayr}}]{semenov03}
{Semenov}, D., {Henning}, T., {Helling}, C., {Ilgner}, M., \& {Sedlmayr}, E.
  2003, \aap, 410, 611

\bibitem[{{Stevenson}(1982)}]{stevenson82}
{Stevenson}, D.~J. 1982, \planss, 30, 755

\bibitem[{{Takahashi}(2001)}]{takahashi01}
{Takahashi}, J. 2001, \apj, 561, 254

\bibitem[{{Thompson}(2006)}]{thompson06}
{Thompson}, M.~J. 2006, {An introduction to astrophysical fluid dynamics}

\bibitem[{{Turner} {et~al.}(2010){Turner}, {Carballido}, \& {Sano}}]{turner10}
{Turner}, N.~J., {Carballido}, A., \& {Sano}, T. 2010, \apj, 708, 188

\bibitem[{{Vardya}(1960)}]{vardya60}
{Vardya}, M.~S. 1960, \apjs, 4, 281

\bibitem[{{Walmsley} {et~al.}(2004){Walmsley}, {Flower}, \& {Pineau des
  For{\^e}ts}}]{walmsley04}
{Walmsley}, C.~M., {Flower}, D.~R., \& {Pineau des For{\^e}ts}, G. 2004, \aap,
  418, 1035

\bibitem[{{Wuchterl}(1993)}]{wuchterl93}
{Wuchterl}, G. 1993, Icarus, 106, 323

\end{thebibliography}

\appendix
\section{Equation of State Table Extension}\label{EOStables}

In this study we consider atmosphere growth in the outer parts of protoplanetary disks ($5<a<100$ AU), where temperature and pressure drop to as little as $T \sim 20$ K and $P \sim4\times10^{-6}$ dyn cm$^{-2}$ for our fiducial disk model (see equations \ref{eq:diskb} and \ref{eq:Pd}). We model the nebular gas using the EOS tables of \citet{saumon95}. However, these tables only cover the relatively high temperature and pressure ranges $2.1 < \log_{10} T(\rm{K})<7.06$ and $4.0<\log_{10}P$(dyn cm$^{-2})<19.0$. We thus need to extend the tables to lower $T$ and $P$. We calculate $\delad$ for

\begin{eqnarray}
1.0 & < & \log_{10} T <2.1 \\ 
-5.4& < & \log_{10} P<4.0 
\end{eqnarray} 
using the following method.

%In this section we explain the procedure for extending and interpolating the \cite{saumon95} EOS tables. The EOS takes into account non ideal interactions, and includes physical treatments of dissociation and ionization. However, the \cite{saumon95} EOS tables only cover a relatively high range of temperatures and pressures: $2.10 < \log_{10} T(\rm{K})<7.06$ and $4<\log_{10}P$(dyn cm$^{-2})<19$. We consider cold disks, where the temperature and pressure drop to $\sim 20$ K and $\sim 10^{-4}$ dyn cm$^{-2}$, respectively (see equations (\ref{eq:diskb}) and (\ref{eq:Pd})). It is therefore necessary to extend the \cite{saumon95} EOS tables to lower temperature and pressure values.

%We choose $\log_{10} T (\rm{K})=1$ and $ \log_{10}P$(dyn cm$^{-2})=-4.4$ as our lower boundaries for temperature and pressure, respectively. Our temperature and pressure grid becomes: $1 < \log_{10} T(\rm{K})<7.06$ and $-4.4<\log_{10}P$(dyn cm$^{-2})<19$. The other thermodynamic variables in the tables are calculated as follows.

\subsection{Hydrogen}

\label{hydrogen}

Following \citet{kittel}, we calculate $\delad$ from the partition function for the internal energy of a system of hydrogen gas molecules (see also \citealt{dangelo13} for EOS calculations that take into account hydrogen isomers).  We begin by writing the partition function $Z$ of a gas molecule of mass $m$ as the product of the partition functions associated with each type of internal energy:

%\begin{equation}
%\label{eq:z}
%Z=Z_t Z_r Z_v Z_e Z_n,
%\end{equation}

\begin{equation}
\label{eq:zagain}
Z=Z_t Z_r Z_v \;\;,
\end{equation} 

\noindent where $Z_t$, $Z_r$, $Z_v$ are associated with translation, rotation, and vibration, respectively.\footnote{We ignore electronic and nuclear excitation as they are only important at temperatures much higher than our regime of interest.}  %For hydrogen, electronic and nuclear excitation are only significant at temperatures higher than our region of interest ($\theta_e \approx 12000$ K and $\theta_n >> \theta_e$, where $\theta_e$ and $\theta_n$ are the characteristic temperatures for electronic and nuclear excitation, respectively). As such, we will only take into account the translation, rotation and vibration of the hydrogen molecule:

%In what follows we present and briefly derive expressions for thermodynamic variables based on quantum mechanics principles. More details on the derivations can be found in \citet{kittel}.

In the classical limit, the molecule's center of mass motion generates

\begin{equation}
\label{eq:Zt}
Z_t=(m k_B T/2 \pi \hbar^2)^{3/2} V,
\end{equation}
where  $T$ and $V$ are the gas temperature and volume, respectively, and $\hbar$ is the reduced Planck constant. The rotational partition function is

\begin{equation}
\label{eq:Zr}
Z_r=\sum_{j=0}^\infty (2 j+1) \exp{\Big[\frac{-j (j+1)\Theta_r}{T}\Big]},
\end{equation}

\noindent where the characteristic temperature for rotational motion $\Theta_r \approx 85$ K  for hydrogen. However, molecular hydrogen occurs in two isomeric forms: parahydrogen with a symmetric (even) rotational wavefunction, and orthohydrogen with an antisymmetric (odd) wavefunction (see Section \ref{deladtable}). %Parahydrogen with an even rotational wave function, while orthohydrogen can only have an antisymmetric (odd) wave function. 
The rotational partition functions for ortho- and parahydrogen are thus
\begin{equation}
\label{eq:Zpara}
Z_{\rm{r,para}}=\sum_{j=0}^\infty \frac{1+(-1)^j}{2} (2 j +1) \exp\Big[-\frac{j(j+1)\Theta_r}{T}\Big]
\end{equation}
and
\begin{equation}
\label{eq:Zortho}
Z_{\rm{r,ortho}}=3\sum_{j=0}^\infty \frac{1-(-1)^j}{2} (2 j +1) \exp\Big[-\frac{j(j+1)\Theta_r}{T}\Big] \;\;.
\end{equation}
The factor of 3 in Equation (\ref{eq:Zortho}) accounts for the three-fold degeneracy of the ortho state.

 In thermal equilibrium, the spin isomers have a combined partition function $Z_{\rm r}=Z_{\rm{r, ortho}}+Z_{\rm{r,para}}$, which can be written

\begin{equation}
\label{eq:Zrspin}
Z_r=\sum_{j=0}^\infty (2-(-1)^j) (2j+1) \exp{\Big[\frac{-j (j+1) \Theta_r}{T}\Big]} \;\;.
\end{equation}
For a fixed 3:1 ortho-to-para ratio, the combined partition function is $Z_{\rm r}=Z_{\rm r,para}^{1/4} Z_{\rm r, ortho}^{3/4}$. In our range of temperatures of interest, we find that $Z_r$ converges after about 25 terms in the series, for both the equilibrium and 3:1 cases.

Note that $Z_{\rm r, ortho} \rightarrow 0$ and $Z_{\rm r,para} \rightarrow 1$ as $T \rightarrow 0$. This is inconsistent with a fixed 3:1 ortho-to-para ratio, since $Z_{\rm r, ortho} \rightarrow 0$ implies that there is no orthohydrogen in the system. \citet{boley07} and \citet{dangelo13} ensure that this requirement is not violated by using a normalized orthohydrogen partition function,  $Z'_{\rm r, ortho}=Z_{\rm r, ortho} \exp(2 \theta_{\rm r}/T)$, which reduces the energy of the lowest rotational state for orthohydrogen from $2k_{\rm B}\theta_{\rm r}$ per molecule to zero. This decreases the total internal energy of the system by a constant factor, but does not change $c_{\rm v}$, $\delad$, or the relative internal energies of static atmospheric profiles, and therefore does not affect atmospheric evolution.

Finally, the partition function for vibrational motion is given by:

\begin{equation}
\label{eq:Zv}
Z_v=[1-\exp{(\theta_v/T)}]^{-1},
\end{equation}

\noindent where the characteristic temperature for vibrational motion is $\theta_v  \approx 6140$ K for hydrogen. 

For a system of $N$ particles of mass $m$, the partition function of the ensemble is $Z_{\rm N}=(1/N!)Z^N$. Given $Z_{\rm N}$ as a function of $(V, T)$, the internal energy per mass, entropy per mass and specific heat capacity can be written as

%\begin{equation}
%\label{eq:U}
%U_N=k T^2 \Big(\frac{\partial \log{Z}}{\partial T}\Big)_{V, N}
%\end{equation}
%
%\begin{equation}
%\label{eq:S}
%S_N=k \log{Z} + \frac{U_N}{T}
%\end{equation}
%
%The energy, and entropy per mass and specific heat capacity will subsequently be:

\begin{equation}
\label{eq:u}
u=\mathcal{R}T^2 \Big(\frac{\partial \ln{Z}}{\partial T}\Big)_{V}
\end{equation}
%{\bf CHANGED BIG U TO LITTLE u IN EQUATIONS BELOW: CONFIRM}
\begin{equation}
\label{eq:s}
S=\mathcal{R} \ln{Z} + \frac{u}{T}-\frac{\mathcal{R}}{N} \ln N!
\end{equation}
\begin{equation}
\label{eq:cv}
c_{\rm V}=\Big(\frac{\partial u}{\partial T}\Big)_{V}.
\end{equation}
Note that, following the convention of \citet{saumon95}, we use $S$ to denote entropy per mass ($[S]=$erg K$^{-1}$ g$^{-1}$).  Since $Z=Z_t Z_r Z_v$, we may write $u=u_t+u_r+u_v$ and $S=S_t+S_r+S_v$, where variables subscripted $t$, $r$, and $v$ are the quantities corresponding to the individual translation, rotation and vibration partition functions, respectively. We include the term $\mathcal{R}/N \ln N!$ from Equation (\ref{eq:s}) in $S_{\rm t}$.

%\textbf{Using the translational partition function, internal energy and entropy expressions for an ensemble of $N$ identical particles of mass $m$, i.e. $Z_N=(1/N!) Z_t^N$, $U_N=k_B T^2 (\partial \ln{Z_N}/\partial T)_{V, N}$, and $S_N=k_B \ln{Z_N} + U_N/T$, Stirling's approximation $\ln N! \approx N \ln N-N,$ and the ideal gas law $P=\rho \mathcal{R} T$, the {\bf resulting} entropy per mass due to translational motion can be expressed as:}

 In the temperature regime for which the rotational states of $\rm{H}_2$ are selectively occupied, the total number of particles in the system is constant and we may use the ideal gas law, $P=\rho \mathcal{R} T$. With Equations (\ref{eq:u}), (\ref{eq:s}), and Stirling's approximation, $\ln N! \approx N \ln N-N$, the resulting entropy per mass due to translational motion is

\begin{equation}
\label{eq:st}
S_{\rm t}=\mathcal{R} \Big[ \frac{5}{2} \ln{T} - \ln{P} + \ln \Big( \frac{(2 \pi)^{3/2} \mathcal{R}^{5/2} m^4}{h^3}\Big) +\frac{5}{2} \Big]
\end{equation}
Equation (\ref{eq:st}) is known as the Sackur-Tetrode formula. % \textbf{and is only applicable to an ideal gas REALLY? WE'RE USING IT FOR THE TRANSLATIONAL COMPONENT IN A NON-IDEAL GAS}. 
%\textbf{We can use the ideal gas law to derive Equation (\ref{eq:st}) for the $\rm{H}_2$ molecule because the total number of particles in the system, and hence the mean molecular weight and specific gas constant $\mathcal{R}$, do not change with temperature in the regime where the rotational states are selectively occupied.} 

The internal energy per mass due to translational motion is given by:

\begin{equation}
\label{eq:ut}
u_t=\frac{3}{2} \mathcal{R} T
\end{equation}

Putting all of the above together, we can now evaluate the thermodynamic quantities needed to extend the \cite{saumon95} EOS tables to low temperatures and pressures.

\begin{figure}[h!]
\centering
\includegraphics[scale=.8]{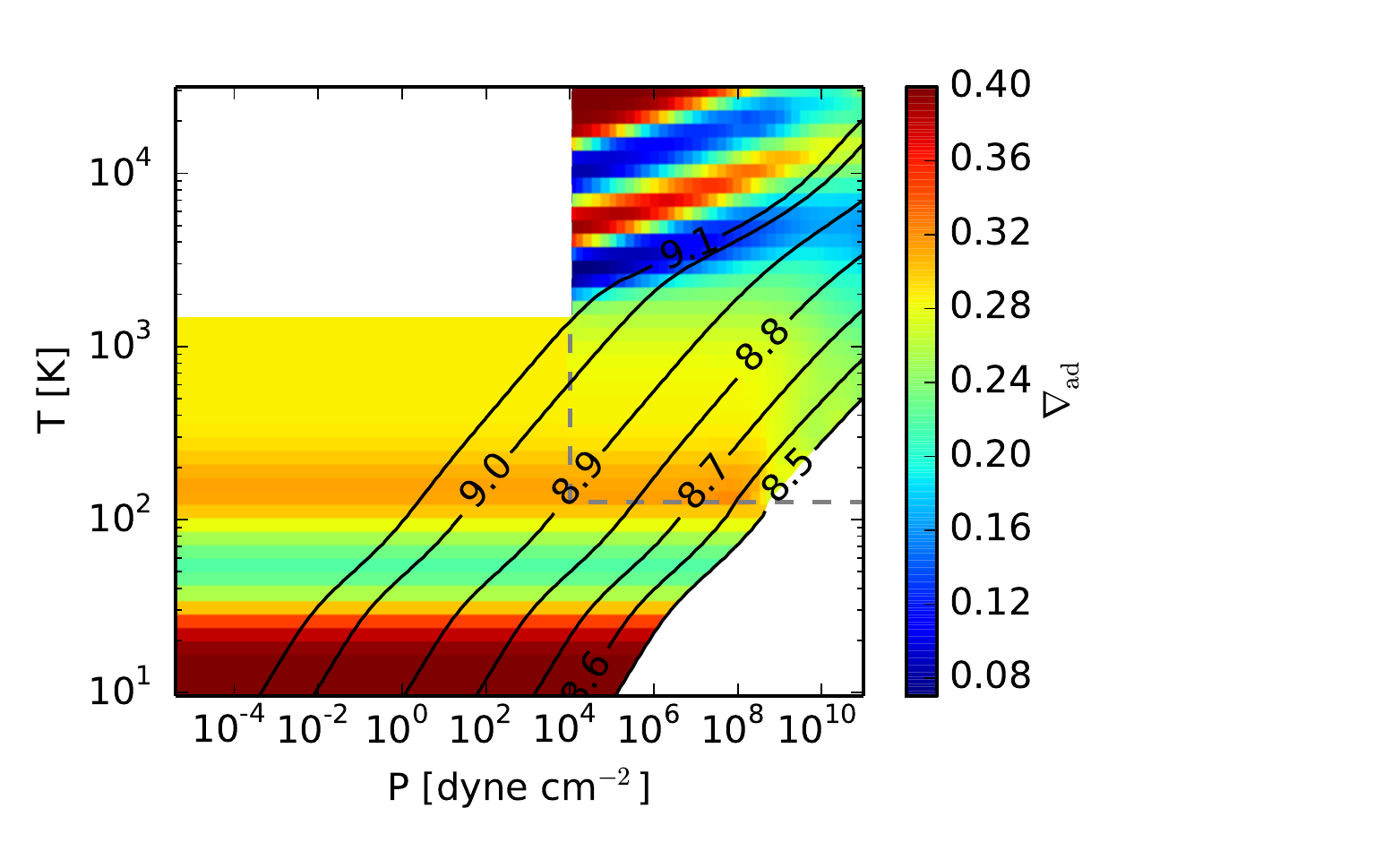}
\caption{Contour plot of the hydrogen adiabatic gradient $\delad$ as a function of gas temperature and pressure. The upper right rectangle encloses the region described by the original \citet{saumon95} EOS tables, while the rest of the plot is our extension to lower temperatures and pressures for an equilibrium mixture of ortho- and parahydrogen. The black curves represent constant entropy adiabats with labels $\log_{10}(S)$, where $S$ [erg K$^{-1}$ g$^{-1}$] is the absolute entropy per unit mass.  At high temperatures, hydrogen dissociates and ionizes, while at low temperatures the rotational states of the hydrogen molecule are only partially excited and it no longer behaves like an ideal diatomic gas. Regions in which the EOS is invalid or has not been computed are masked in white (see text).}
\label{fig:deladH}
\end{figure}

\begin{enumerate}

\item{\textbf{Density.}} In the low temperature, low pressure regime, $\rho$ is related to $T$ and $P$ by the ideal gas law. %hydrogen is molecular and behaves like an ideal gas. As such, the density in this region follows the ideal gas law $P=\rho \mathcal{R} T$.
\item{\textbf{Internal energy per mass.}} $u=u_t+u_r+u_v$, where $u_t$ is given by Equation (\ref{eq:ut}), and $u_r$ and $u_v$ are determined using Equations (\ref{eq:Zrspin}), (\ref{eq:Zv}) and (\ref{eq:u}) above.
\item{\textbf{Entropy per unit mass}}. Similarly, $S=S_t+S_r+S_v$, where $S_t$ is given by Equation (\ref{eq:st}), and $S_r$ and $S_v$ can be determined from Equation (\ref{eq:s}) and the calculated expressions for $u_r$ and $u_v$, respectively.
\item{\textbf{Entropy logarithmic derivatives}}. The logarithmic derivatives $S_T$ and $S_P$ are given by:

\begin{equation}
\label{eq:sT}
S_T=\frac{\partial \ln{S}}{\partial \ln{T}} \Big |_P
\end{equation}
and
\begin{equation}
\label{eq:sP}
S_P=\frac{\partial \ln{S}}{\partial \ln{P}} \Big |_T
\end{equation}
We calculate $S_T$ and $S_P$ through finite differencing. 

\item{\textbf{Adiabatic gradient $\delad$}}. The adiabatic gradient is defined as:

\begin{equation}
\label{eq:deladSP}
\delad=\frac{\partial \ln{T}}{\partial \ln{P}} \Big |_S = -\frac{S_P}{S_T}
\end{equation}

We evaluate $\delad$ from the tabulated values for $S_T$ and $S_P$ determined above. Figure \ref{fig:deladH} shows a contour plot of $\delad$ as a function of temperature and pressure for the extended EOS table for hydrogen, assuming thermal equilibrium between the spin isomers. For the 3:1 ortho-to-para ratio, $\delad$ decreases continuously with $T$ for $T \lesssim 200$ K, in contrast with the equilibrium case, in which $\delad$ sharply decreases, then increases as $T$ goes down. Our extension is only valid for $T \lesssim 2000$ K, since it does not take into account hydrogen dissociation. We choose $T=1500$ K as a conservative temperature cutoff. While we account for vibrational motion for completeness, its contribution is negligible in the temperature regime of interest. \citet{saumon95} do not compute the EOS at very high pressures, since hydrogen is solid or may form a Coulomb lattice in this regime, and thus their EOS treatment is no longer valid. While the boundaries of the region in which the free-energy EOS treatment fails can be determined from fundamental thermodynamic constraints, such calculations are not the object of this work. Instead, we choose as boundary a constant entropy curve ($\log(S)=8.4$) above the region in which the \citet{saumon95} model fails. The expressions derived above are sufficient to give good results for the colored regions of the extended map, which fully cover the temperature and pressures ranges required by our models.

 %Our extension smoothly matches the original tables for $8.80<\log{S}$(K g$^{-1})<9.07$ (\textbf{numbers are wrong, change once you have the final figure version}).

\end{enumerate}

\subsection{Helium}

\begin{figure}[h!]
\centering
\includegraphics[scale=.8]{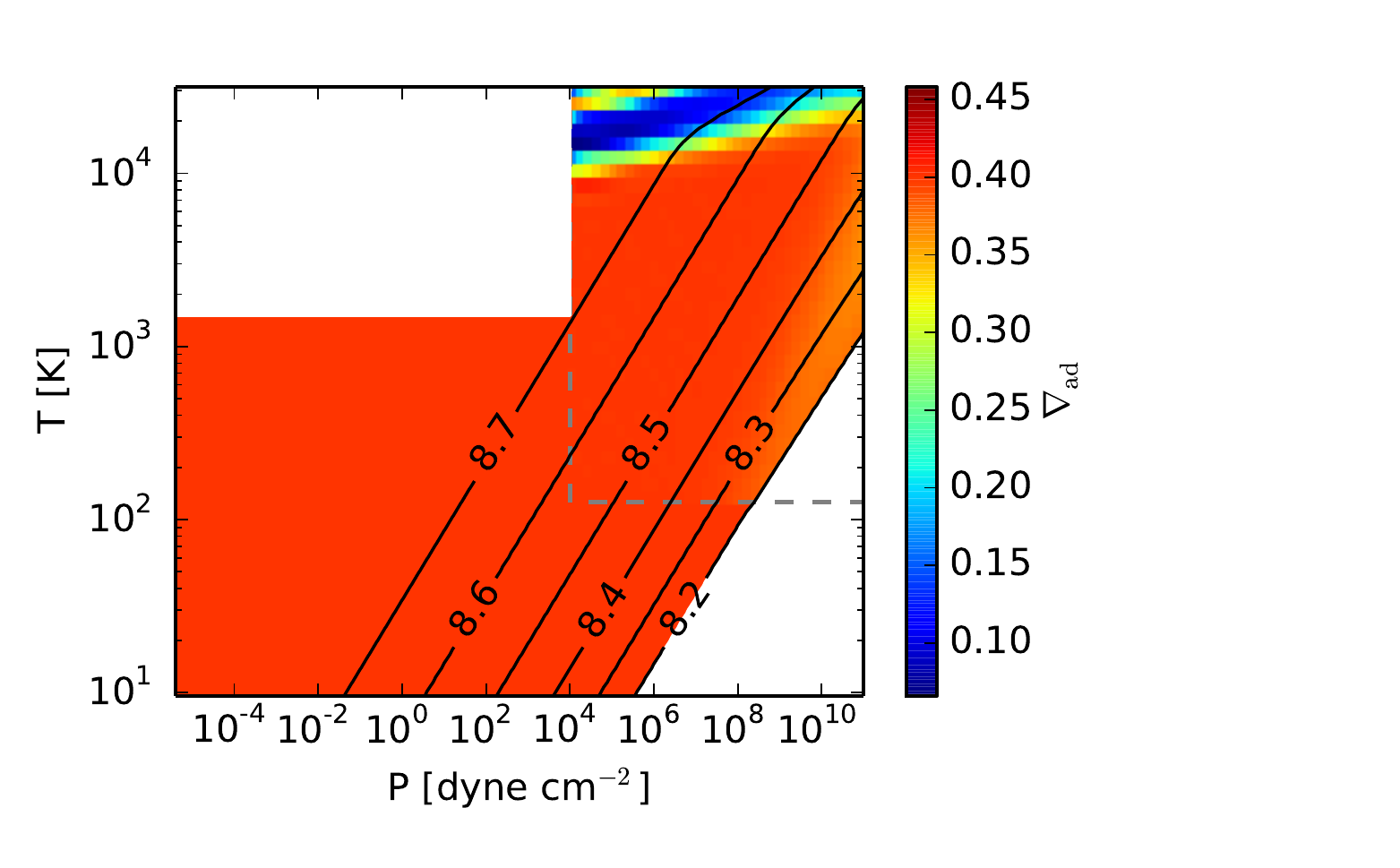}
\caption{Same as Figure \ref{fig:deladH} but for pure helium. Helium ionizes at $T \gtrsim 10,000$ K, but behaves as an ideal monatomic gas otherwise. We choose $T=7,000$ K as a conservative temperature cutoff above which our  extension is no longer valid (masked in white). The EOS has not been computed in the lower-right region of the plot (see text).}
\label{fig:deladHe}
\end{figure}

We extend the helium EOS tables based on a similar procedure. Since helium is primarily neutral and atomic at low temperatures and pressures, we treat it as an ideal monoatomic gas and  only take into account the translational component of the partition function (\ref{eq:Zt}). Figure \ref{fig:deladHe} shows $\delad$ as a function of temperature and pressure for the extended EOS table. %The original and extended table join smoothly for entropy curves between $8.29<\log{S}$(K g$^{-1})<8.77$ in this case (\textbf{numbers are wrong, change once you have the final figure version}).

\vspace{0.2in}

Lastly, we combine Figures \ref{fig:deladH} and \ref{fig:deladHe} to obtain EOS tables for a hydrogen-helium mixture using the procedure described in \citet{saumon95}.  Figure \ref{fig:deladmap} displays results for helium mass fraction $Y=0.3$.

%Using equation (\ref{eq:upartition}) we therefore recover the standard result $U_{\rm r}=\mathcal{R} T$ (refs). Furthermore, we know that the internal energy and entropy per unit mass associated with translation are given by $U_{\rm t}=\frac{3}{2} \mathcal{R}$ and $C_{\rm{v,t}}=\frac{3}{2}\mathcal{R}$, respectively, and so we are able to calculate the total internal energy and specific heat of a diatomic molecule as a function of temperature. An example of the variation of heat capacity with temperature is shown in \citet{kittel}, chapter 3. 

%The partition function associated with rotation is generally written as:
%
%\begin{equation}
%\label{eq:Zr}
%Z_{\rm r}=\sum_0^\infty (2 j +1) \exp\Big[-\frac{j(j+1)\Theta_r}{T}\Big],
%\end{equation}
%with $j$ the angular momentum quantum number \citep{kittel}. Various thermodynamic quantities can be derived from the partition function. For example, the internal energy per unit mass can be written as:
%
%\begin{equation}
%\label{eq:upartition}
%u_{\rm r}=\mathcal{R}T^2 \frac{\partial \log Z}{\partial T}
%\end{equation}
%
%The specific heat at constant volume then easily follows as
%
%\begin{equation}
%\label{eq:cvpartition}
%c_{\rm{v,r}}=\Big(\frac{\partial u}{\partial T}\Big)_V
%\end{equation} 

\section{Adiabatic Gradient Variations}
\label{alldelad}

\subsection{Adiabatic Gradient during Partial Dissociation}\label{deladdiss}

%The adiabatic gradient during dissociation or ionization is a function of temperature and the dissociation or ionization fraction $x$. The Saha equation (e.g., \citealt{kippenhahn90}) relates $x$ to the gas 

The total internal energy of a partially dissociated gas includes contributions from the individual internal energies of the molecules and atoms, as well as from the dissociation energy. The dissociation energy depends on the dissociation fraction $x$ (i.e., the fraction of molecules that have dissociated), which can be found from the Saha equation (see e.g., \citealt{kippenhahn90}, Chapter 14) as a function of temperature and density,

\begin{equation}
\label{eq:saha}
\frac{x^2}{1-x} \propto \frac{T^{3/2}}{\rho} e^{-\chi/k_B T},
\end{equation} 
where $\chi$=4.48 eV is the dissociation energy for molecular hydrogen \citep{blanksby03}.

\begin{figure}[h]
\centering
\includegraphics[width=0.5\textwidth]{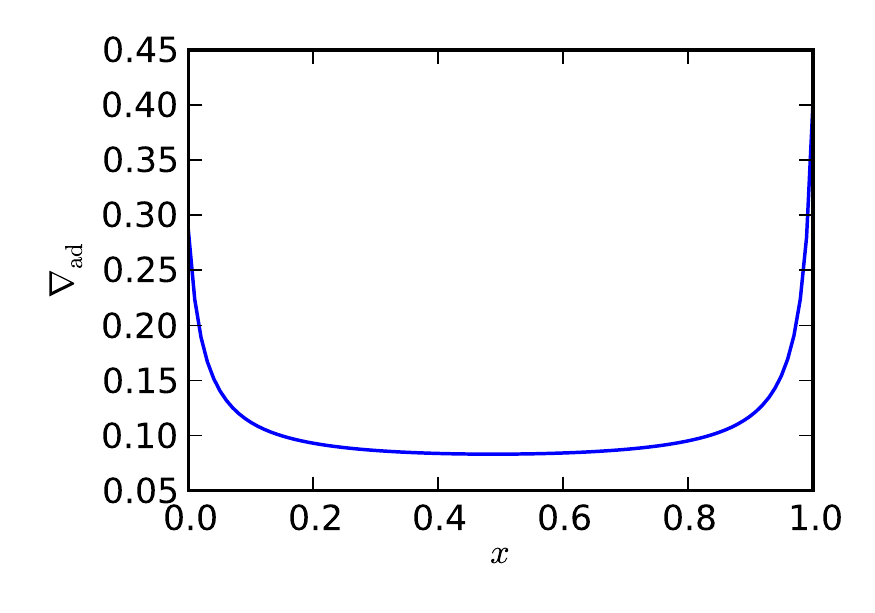}
%%\vspace{-0.5in}
\caption{Adiabatic gradient as a function of the hydrogen dissociation fraction $x$. The adiabatic gradient is $\delad=2/7$ for pure molecular hydrogen ($x=0$) and $\delad=2/5$ for fully atomic hydrogen ($x=1$), and drops to low values during partial dissociation.}
\label{fig:deladdiss}
\end{figure}

The above also holds true for ionization, with the dissociation energy replaced by ionization energy $\chi=13.6$ eV for atomic hydrogen \citep{mandl89}. From the Saha equation one can find an expression for $\rho$ as a function of $T$ and $x$, then derive the adiabatic gradient directly from its definition (Equation \ref{eq:delad}), taking into account the fact that the mean molecular weight in the ideal gas law varies with $x$, hence the pressure will not only be a function of $T$ and $\rho$ but also of $x$ (see \citealt{kippenhahn90}, Chapter 14.3 for a detailed derivation). The final expression for the adiabatic gradient during ionization is 

\begin{equation}
\label{eq:deladioniz}
\delad=\frac{2+x (1-x) \Phi_H}{5+x(1-x)\Phi_H^2},
\end{equation}
with $\Phi_H \equiv \frac{5}{2}+\frac{\chi}{k_B T}$. The derivation of $\delad$ during dissociation is more involved mathematically (see, e.g., \citealt{vardya60}) and leads to a slightly more complicated final expression,

%One can derive a simple expression for the adiabatic gradient as a function of $x$ and $T$ from it's definition (equation \ref{eq:delad}), using the 

%The adiabatic gradient follows as (see \citealt{vardya60} for a detailed derivation)

\begin{equation}
\label{eq:deladdiss}
\delad=\frac{1+x+ \frac{x(1-x^2)}{2} \frac{\chi}{k_B T}}{5 x + \frac{7(1-x)}{2} + \frac{x(1-x^2)}{2} \Big(\frac{\chi}{k_B T}\Big)^2} \, .
\end{equation} 
Using Equation (\ref{eq:deladdiss}), we recover $\delad=2/7$ for $x=0$ (no ongoing dissociation hence hydrogen is purely molecular and diatomic) and $\delad=2/5$ for $x=1$ (hydrogen is fully dissociated into atoms and hence monatomic). Figure \ref{fig:deladdiss} shows the dependence of $\delad$ on the dissociation fraction, for $T=3000$ K, the temperature at which dissociation typically occurs \citep{langmuir12}. The adiabatic gradient drops substantially during partial dissociation, since part of the internal energy is used in dissociation rather than in increasing the temperature of the system.

\subsection{Adiabatic Gradient during Conversion of Spin Isomers}
\label{deladspin}

The adiabatic gradient scales as $\delad \sim 1/c_{\rm V}$.
%  where $c_{\rm V} \sim c_{\rm V,t}+c_{\rm V,r}$.
% is the specific heat capacity at constant volume and $c_{\rm V, t}$ and $c_{\rm V,r}$ are the translational and rotational components, respectively.  
The translational component of the heat capacity, $c_{\rm V, t}=3\mathcal{R}/2$, is independent of temperature.  As $c_{\rm V,r}$ increases, $\delad$ declines. We can therefore understand how conversion between spin isomers affects $\delad$ by studying the dependence on temperature of $c_{\rm V, r}$ of the ortho-para mixture. 

\begin{figure}[h]
\centering
\includegraphics[width=0.5\textwidth]{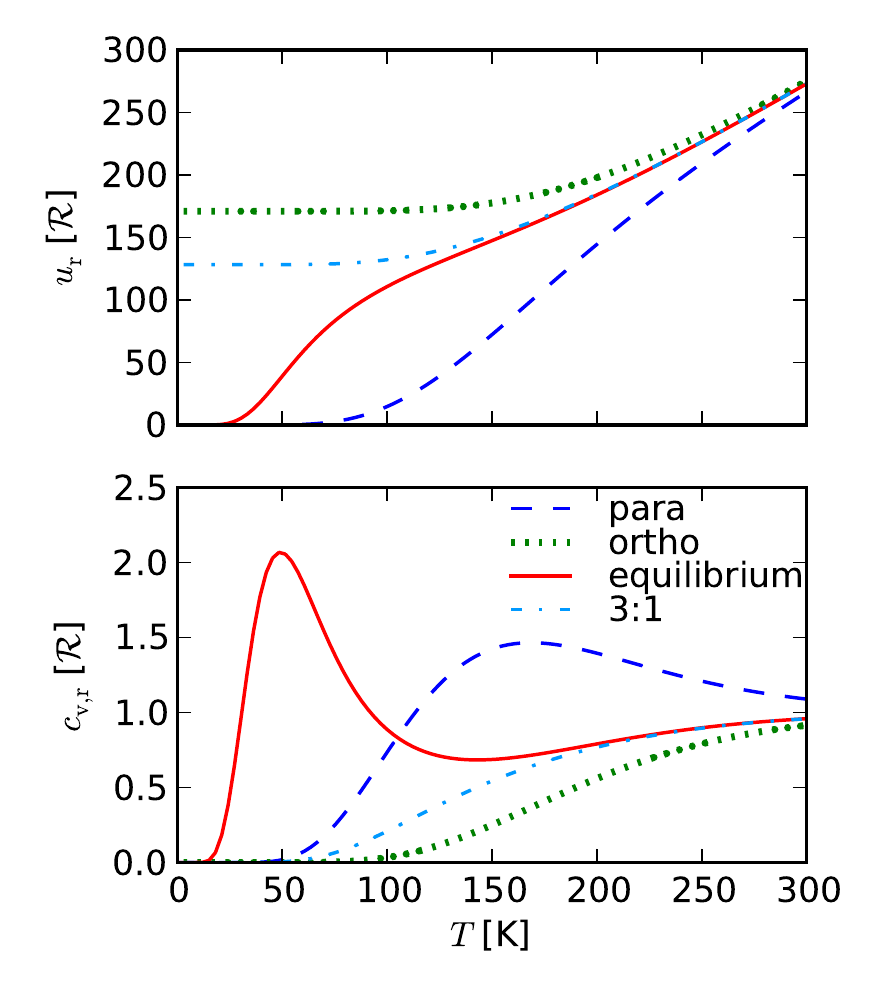}
%%\vspace{-0.5in}
\caption{Internal energy per unit mass and specific heat capacity associated with rotation for parahydrogen (dashed blue), orthohydrogen (dotted green), the equilibrium mixture (solid red) and a fixed 3:1 ortho-to-para ratio (dash-dotted light blue) as a function of temperature. After \citet{farkas35}, Figure 1.}
\label{fig:ucvr}
\end{figure}

The internal energy per unit mass and specific heat capacity associated with rotation for the individual isomers, for the equilibrium mixture, and for a fixed ortho-para ratio of 3:1 can be derived from their partition functions (see \App{EOStables} for details), and are plotted in Figure  \ref{fig:ucvr} (after \citealt{farkas35}, Figure 1). At low temperatures, parahydrogen is in the $j=0$ state and has no rotational energy, while orthohydrogen is in the $j=1$ state and has the energy of its first rotational level. Both para- and orthohydrogen, as well as their equilibrium mixture, behave like monatomic gases at low temperatures and thus have zero rotational heat capacity. This is consistent with $\delad=2/5$ at low temperatures as seen in Figure \ref{fig:deladmap}. As the temperature increases, the energetically higher-lying rotational states of para- and orthohydrogen are populated and the heat capacity of both spin isomers increases as a result. We note that the heat capacity of the equilibrium mixture is not a weighted average of the heat capacities of the individual components because it takes into account both the rotational energy uptake of para- and orthohydrogen, and also the shift in their equilibrium concentrations with temperature. This results in a peak in the heat capacity of the mixture around $\sim$$50$ K, as seen in the bottom plot of Figure \ref{fig:ucvr}. It follows that the adiabatic gradient has to reduce, reach a minimum, then increase as the temperature rises, as shown in Figure \ref{fig:deladmap}. In contrast, the heat capacity for the 3:1 ortho-para ratio mixture will be a weighted average between the individual ortho- and para- components, and will hence have intermediate values between the two, as displayed in Figure \ref{fig:ucvr}.

\begin{figure}[h]
\centering
\includegraphics[width=0.5\textwidth]{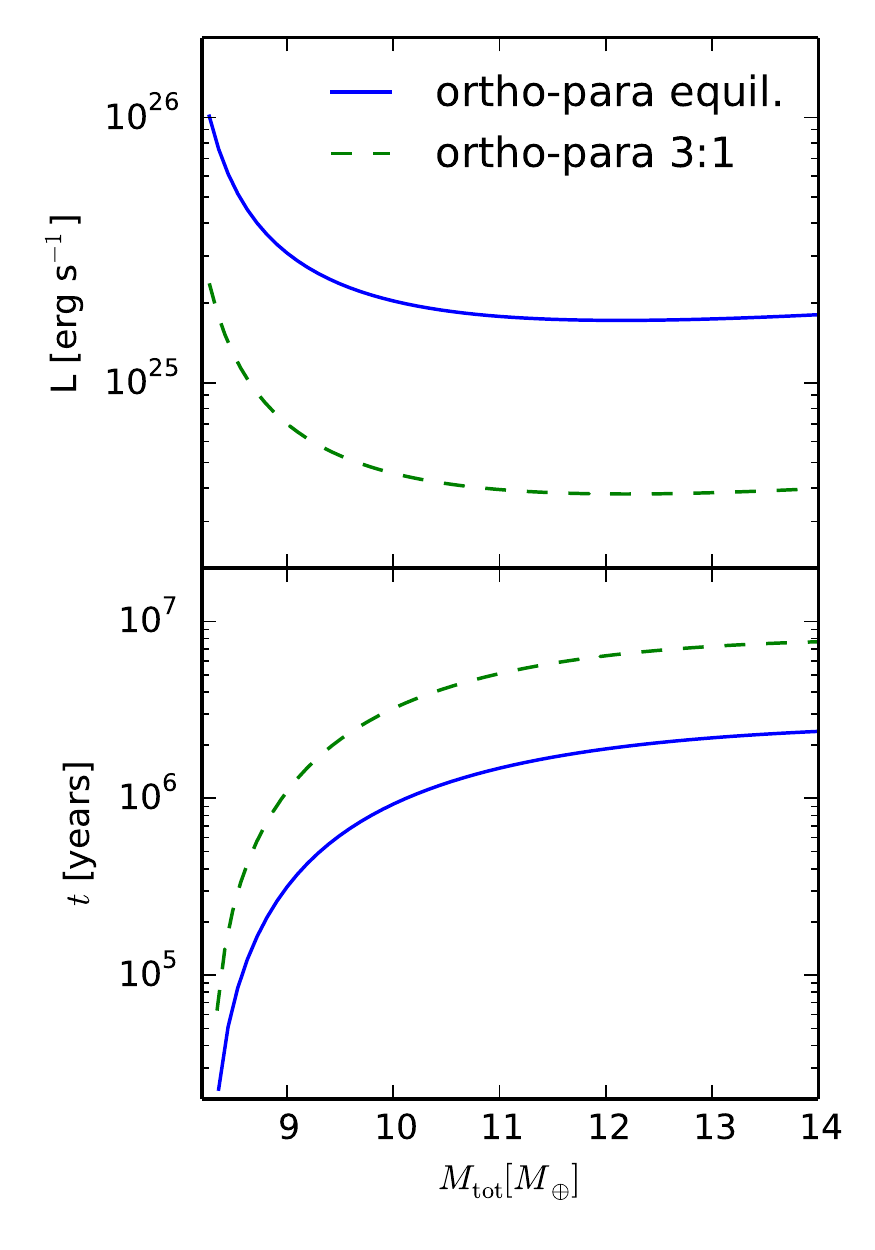}
%%\vspace{-0.5in}
\caption{Evolution of the luminosity and elapsed time during atmospheric growth around a $8 M_{\oplus}$ core at 100 AU, for a realistic EOS with hydrogen spin isomers in thermal equilibrium (solid line), and with a fixed ortho-to-para ratio 3:1 (dashed line). The assumption of a fixed ortho-to-para ratio increases the runaway accretion time $t_{\rm run}$ by a factor of $\sim$$3$ compared to the equilibrium mixture.}o
\label{fig:Lt_31}
\end{figure}

Figure \ref{fig:Lt_31}, bottom panel, shows that the atmospheric growth time may increase by a factor of $\sim$$3$ if a fixed 3:1 ortho-para ratio is assumed instead of thermal equilibrium between the hydrogen spin isomers.  This enhanced growth time  increases $M_{\rm crit}$.

\section{Grain growth opacity and radiative windows} \label{radwindow}

\begin{figure}[h!]
\centering
\includegraphics[width=0.5\textwidth]{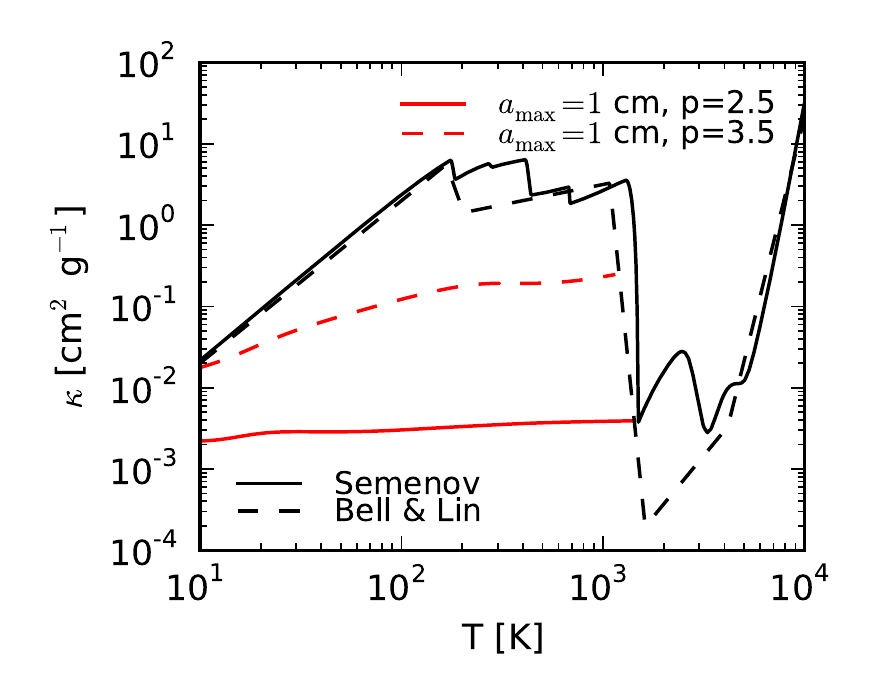}
%%\vspace{-0.5in}
\caption{Rosseland mean opacity of dust grains as a function of temperature for different opacity assumptions. The dashed black curve shows the \citet{bell94} analytic ISM opacity for $\rho=10^{-8}$ g cm$^{-3}$. The solid black curve shows the tabulated opacity of \citet{semenov03} for a dust composition of 'normal' silicates. The dashed red curve shows the \citet{dalessio01} opacity, which takes grain growth into account, for a maximum particle size of 1 cm and a standard collisional cascade grain size distribution ($p=3.5$). The solid red curve is the same as the dashed red curve, but it accounts for coagulation ($p=2.5$).}
\label{fig:opacity}
\end{figure}

The opacity of the interstellar medium is reasonably well constrained and approximate analytic expressions for the Rosseland mean opacity as a function of temperature and density are derived in \citet{bell94}. For low temperatures ($T \lesssim 100$ K) at which ice grains are present, opacity scales with temperature as $\kappa \sim T^2$. Sublimation of ice grains at $\sim$$150$ K and of metal grains at $\sim$$1000$ K results in sharp opacity drops. This is shown in Figure \ref{fig:opacity} for a gas density $\rho=10^{-8}$ g cm$^{-3}$, which is typical for the outer regions of protoplanetary disks. \citet{semenov03} calculate Rosseland mean opacities in protoplanetary disks for grains of different sizes and structure. As shown in Figure \ref{fig:opacity}, their results are in good agreement with \citet{bell94}. However, \citet{semenov03} do not take grain growth into account, which is likely to occur in protoplanetary disks, particularly at the late times when cores form. \citet{dalessio01} compute wavelength dependent opacities for a range of maximum particle sizes and different size distributions. Figure \ref{fig:opacity} shows the integrated Rosseland mean opacity for a maximum particle size of 1 cm and a power law differential size distribution $dN/ds \propto s^{-p}$, with $s$ the grain size and $p=3.5$ for a standard collisional cascade and $p=2.5$ when coagulation is taken into account. We see in Figure \ref{fig:opacity} that this yields a mean opacity that is both lower and less sensitive to temperature, when compared to \citet{bell94} or \citet{semenov03}. However, \citet{dalessio01} only computes opacities for temperatures less than the dust sublimation temperature, appropriate for current observations of dust in protoplanetary disks. As we see in Figure \ref{fig:opacity}, the opacity dramatically decreases during dust sublimation. We thus use the \citet{bell94} opacities for $T \gtrsim 1000$ K, ensuring that they smoothly match the \citet{dalessio01} opacities for lower temperatures.

\begin{figure}[h!]
\centering
\includegraphics[width=0.5\textwidth]{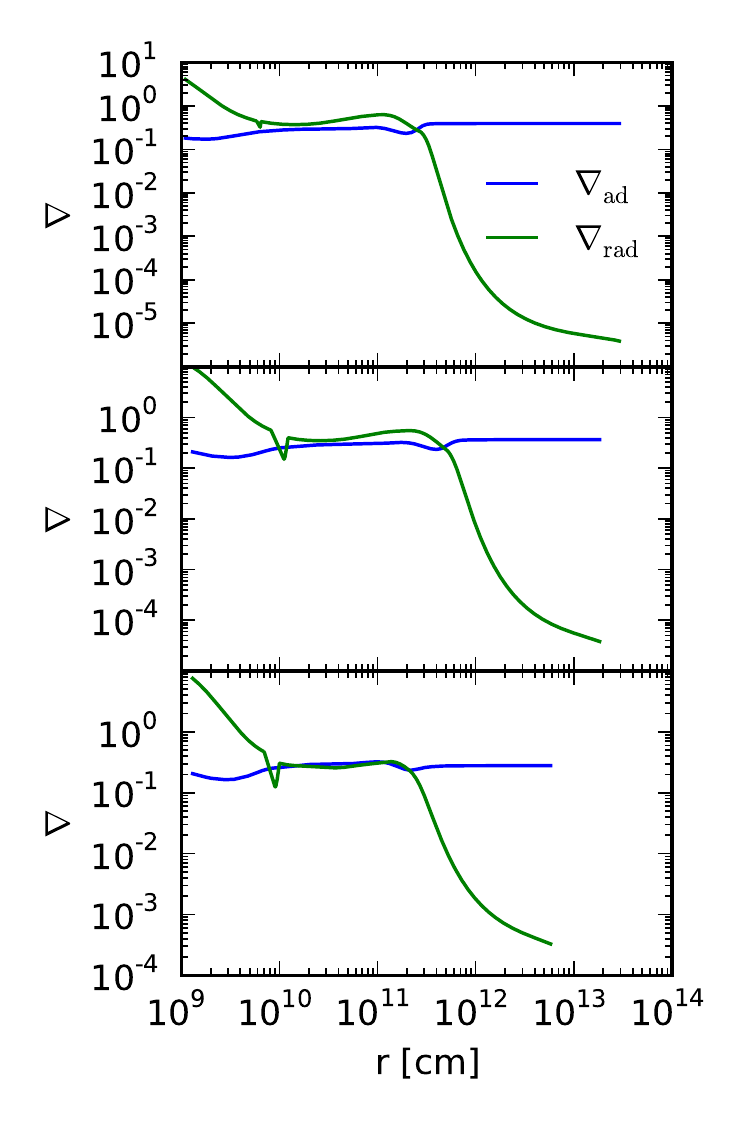}
%%\vspace{-0.5in}
\caption{Snapshots of the radiative and adiabatic gradient as a function of the radial coordinate, for planets with different core masses forming at various locations in the disk. The nebular gas is described by a realistic EOS, with a standard collisional cascade size distribution ($p=3.5$). The sharp drop in opacity due to dust sublimation may generate one or more radiative windows. Top panel: no radiative window for $a=100$ AU and $M\co=3 M_{\oplus}$. Middle panel: the sharp opacity decrease produces one radiative window for $a=50$ AU and $M\co=5 M_{\oplus}$. Bottom panel: the decrease in opacity results in two radiative windows for $a=20$ AU and $M\co=3 M_{\oplus}$.}
\label{fig:delvsr}
\end{figure}

The significant opacity drop due to the sublimation of ice and metal grains lowers the radiative temperature gradient $\delrad$, which may result in one or more inner radiative layers inside the atmosphere of a protoplanet. This is displayed in Figure \ref{fig:delvsr}: depending on the semimajor axis and core mass, the opacity drop will generate no radiative window (top panel), one radiative window (middle panel), or two radiative windows (bottom panel). 

%This is not a problem, however, if most atmospheric luminosity is generated in the innermost convective region of the envelope (see \S\ref{sec:opacity}). We can check this \textit{a posteriori} by using the local energy equation,
%\begin{equation}
%\label{eq:localen}
%\frac{\partial L}{\partial m}=-T \frac{\partial S}{\partial t},
%\end{equation}
%assuming that $\partial S/\partial t$ is constant for a given atmospheric profile. Our model is then valid if $\Delta I \ll I$, where
%\begin{equation}
%\label{eq:I}
%I = \int_{M_{\rm c}}^{M_{\rm{RCB_1}}} T d m
%\end{equation}
%and
%\begin{equation}
%\label{eq:I}
%\Delta I = \int_{M_{\rm{RCB_1}}}^{M_{\rm p}}T d m,
%\end{equation}
%with $\rm{RCB_1}$ the innermost RCB and $M_{\rm p}$ the total planet mass. We find $\Delta I /I \lesssim 30\%$ for all our models of interest (see, however, \S\ref{sec:opacity} and \S\ref{critical} for exceptions). 

%\textit{Radiative windows could, in principle, change atmospheric structure and thus affect evolution if most luminosity is not generated in the innermost convective region of the envelope. } 

{Our idealization of a constant $L$ with radius may be challenged by the presence of radiative windows. While the structure of convective regions is unaffected by $L$, the structure of radiative windows depends on $L$. Fortunately, the assumption of constant luminosity remains reasonable if most of the luminosity is generated in the innermost convective region of the envelope. We can check whether this is true \textit{a posteriori} by using the local energy equation,
\begin{equation}
\label{eq:localen}
\frac{\partial L}{\partial m}=-T \frac{\partial S}{\partial t},
\end{equation}
and integrating it between $M\co$ and $M_{\rm{RCB_1}}$, where $\rm{RCB_1}$ is the innermost RCB. This luminosity can be as little as half of the assumed fixed atmospheric luminosity $L$.

Self-consistently calculating $L(r)$ is not feasible for our code. Instead, we note that since  $\partial S/\partial t$ is fixed in convective regions \citep{arras06}, the luminosity profile is more centrally concentrated than $L \propto m$, which can be implemented simply.  We calculated example profiles using $L \propto m$ and found that though this luminosity scaling may move the location of the outermost RCB, it does not substantially change the luminosity emerging at the top of the atmosphere or the time evolution. %This result may be understood by noting that small changes in the relative normalization of $\delad$ and $\delrad$ in Figure \ref{fig:delvsr} can change the location of the RCB without affecting structure significantly.

\end{document}